%% file: main.tex
\documentclass[aps,pra,twocolumn,superscriptaddress,floatfix,longbibliography, nofootinbib]{revtex4-2}
\usepackage{tabularx}
\usepackage{multirow}
\pdfoutput=1
\input{allcommands.tex}

\input{figures.tex}

\begin{document}

\title{Efficient Circuit Transpilation of Commuting Gates on 2D Grids}

\author{Sabina Dr\u{a}goi}
\affiliation{Institute for Theoretical Physics, ETH Z\"{u}rich, Wolfgang-Pauli-Str. 27, 8093 Z\"{u}rich, Switzerland}
\affiliation{IBM Research, S\"{a}umerstrasse 4, CH-8803 R\"{u}schlikon, Switzerland}

\author{Daniel J. Egger}
\affiliation{IBM Research, S\"{a}umerstrasse 4, CH-8803 R\"{u}schlikon, Switzerland}

\begin{abstract}
Combinatorial optimization problems are central to many applications but can be challenging to solve. 
Quantum approaches such as the Quantum Approximate Optimization Algorithm (QAOA) offer new tools with which to tackle such problems. However, QAOA circuits inherit the interaction structure of the target Hamiltonian, often resulting in deep circuits when compiled onto hardware with limited connectivity. Efficient transpilation is therefore critical to their practical performance.
In this work, we propose a transpilation scheme for circuits consisting of blocks of commuting two-qubit gates on two-dimensional lattices. Unlike standard approaches based on random initial mappings and fixed routing, our method alternates between constructing problem-dependent SWAP-layer sequences and updating the qubit layout. By adapting the routing to the required interactions, this yields significantly shorter circuits for graphs with few edges.
We benchmark our approach on QAOA instances for Maximum Cut (MC) on Random Regular graphs and Maximum Independent Set (MIS) on Erdős–Rényi graphs. Compared to standard methods, we reduce circuit depth and gate count by about a factor of two, enabling experiments with up to $80$ qubits and improving approximation ratios by up to $6.6\%$ for MC and $9.3\%$ for MIS.
\end{abstract}

\maketitle

\section{\label{sec:intro}Introduction}
Combinatorial optimization problems, such as routing~\cite{toth2014vehicle}, scheduling~\cite{pinedo2016scheduling}, and resource allocation~\cite{katoh1998resource}, are ubiquitous across science and industry. Many of these problems are NP-complete, meaning that while candidate solutions can be verified efficiently, finding optimal solutions is believed to require super-polynomial time in the worst case~\cite{garey1979computers}. Their difficulty stems from the discrete nature of the solution space, which precludes the use of gradient-based methods and instead requires exploring a combinatorially large set of configurations~\cite{papadimitriou1998combinatorial,nemhauser1988integer}. 
As a result, we often rely on heuristics or approximation techniques whose performance deteriorates with increasing problem size~\cite{williamson2011approximation, vazirani2001approximation, feige1998threshold}, motivating the search for alternative computational paradigms~\cite{montanaro2016quantum, bharti2022noisy, zhao2026exponential, hangleiter2026has}.

Quantum computing offers a promising approach to combinatorial optimization by mapping such problems to Hamiltonians whose ground states represent optimal solutions~\cite{lucas2014ising}. 
Quantum algorithms can then leverage superposition, entanglement and interference to prepare states that approximate these target ground states.
The Quantum Approximate Optimization Algorithm (QAOA)~\cite{farhi2014quantum} constructs an Ansatz with $p$ alternating applications of time-evolutions of a problem Hamiltonian $H_C$ and a mixing Hamiltonian $H_M$. 
The duration of each time-evolution is chosen to prepare low-energy states of $H_C$.
In the limit $p \to \infty$, the adiabatic theorem guarantees the existence of parameters that prepare the ground state of $H_C$ when initializing the quantum system in the ground state of $H_M$~\cite{rajak2023quantum, finnila1994quantum}. 
In practice, however, QAOA must operate at small $p$ values due to hardware noise which limits circuit depth. 
The problem Hamiltonian $H_C$ typically induces interactions between many pairs of qubits, which are implemented on digital quantum computers using two-qubit gates. 
Hardware with limited connectivity must thus route qubits into adjacent positions with SWAP gates. 
This routing increases circuit depth and accumulated error. 
To mitigate this, we can engineer QAOA circuits generated from sparsified cost Hamiltonians~\cite{druagoi2026approximate, wang2025chain}, which may not always be desirable since we lose the adiabatic convergence guarantee.

Transpilation is the process of converting a high-level quantum circuit into a logically-equivalent one that the hardware can interpret. 
Finding an optimal qubit routing for general circuits is itself an NP-complete problem~\cite{zou2024lightsabre, li2019mapping, MaslovCircuitPlacement}. However, this task can be simplified for circuits composed of blocks of commuting two-qubit gates, such as the QAOA Ansatz. 
Current state-of-the-art transpilers leverage commutativity to design a pre-determined network of SWAP gates, called a SWAP strategy, to systematically generate all pairwise interactions~\cite{weidenfeller2202scaling, matsuo2023sat}. 
This approach is practical on simple hardware connectivities, such as a line of qubits.
However, designing SWAP strategies for higher connectivities is a much harder task. 
Furthermore, pre-determined SWAP networks can be suboptimal for
cost Hamiltonians with few interactions.

The physical qubit connectivity of cross-resonance superconducting qubit hardware was historically very sparse to avoid frequency collisions and cross-talk~\cite{rigetti2010cr, chamberland2020topological}. 
Recent hardware now supports two-dimensional grid connectivity, a step toward the richer connectivity required by new error correction codes~\cite{bravyi2024high}.
Here, we exploit this grid connectivity to improve the routing of sparse optimization problems.

The rest of this paper is structured as follows.
We introduce two benchmark problems and their corresponding QAOA circuits in Sec.~\ref{sec:qaoa}. 
In Sec.~\ref{sec:swap_strategies} we review the state-of-the-art transpilation method for linear qubit arrays. 
In Sec.~\ref{sec:swaps_for_grids}, we study SWAP strategies for grids and show how to tailor the SWAP network to the target problem.
Sec.~\ref{sec:simulations} shows the advantage of our method over standard approaches on the two benchmark problems. 
Sec.~\ref{sec:hardware} then demonstrates the resulting improved noise robustness and overall performance on quantum hardware. 
We discuss our results and conclude in Sec.~\ref{sec:conclusion}.
 
\section{\label{sec:qaoa_and_problems}Background}
\subsection{\label{sec:qaoa}Benchmark problems and QAOA}

We first consider the Maximum Cut (MC) problem. 
It is an NP-complete unconstrained combinatorial optimization problem defined over $n$ binary decision variables $x \in \{ 0,1\}^n$ and often used to study QAOA performance~\cite{koch2025decathlon, zhou2020quantum, brandhofer2022benchmarking, willsch2020benchmarking}. The goal is to split a set $V$ of $n$ nodes of a graph $G = (V,E)$ in two such that the sum of the edges connecting the two subsets is maximized. Formally, a decision variable $x_i$ is assigned to each node $i \in V$ to indicate which side of the cut the node is on. 
Solving MC requires finding the binary string $x$ that maximizes 
\begin{equation}\label{eq:cost_function_MC}
    f_{\text{MC}}(x) = \sum_{(i,j) \in E} \Big( x_i (1-x_j) + x_j (1-x_i) \Big).
\end{equation}

Quantum computers can tackle such problems by mapping $f(x)$ to an Ising Hamiltonian, also known as the cost Hamiltonian $H_C$, whose ground state maximizes $f(x)$. This is done through the change of variables $x_i=\frac{1-z_i}{2}$ and promoting $z_i$ to Pauli $Z$ operators $Z_i$~\cite{lucas2014ising}. Mapping $f_{\text{MC}}(x)$ to a Hamiltonian~\cite{lucas2014ising} results in the cost operator
\begin{equation}\label{eq:maxcut_ham}
    H_{\text{MC}} = \frac{1}{2} \sum_{(i,j) \in E}  Z_i Z_j,
\end{equation}
with a ground state that maps to the maximum cut. 

Similarly, the Maximum Independent Set (MIS) problem is an NP-complete problem 
widely used as a benchmark for quantum strategies~\cite{farhi2014quantum, zhou2020quantum, koch2025decathlon}.
Given a graph $G=(V,E)$, we seek the largest subset $N \subseteq V$ such that no two nodes in $N$ are connected by an edge. This can be formulated as maximizing $\sum_{i \in V}  x_i$, where $x_i \in \{0,1\}$, subject to $x_ix_j = 0, \forall (i,j) \in E$, which we impose via the cost function
\begin{equation}\label{eq:cost_function_MIS}
    f_{\text{MIS}}(x) = \sum_{i \in V} x_i - M \sum_{(i,j) \in E}  x_i x_j
\end{equation}
through a Lagrangian multiplier $M$. 
Choosing $M > 1$ ensures that violating the independence constraint is never advantageous. If two adjacent nodes are both selected, it is enough to deselect just one of them to locally satisfy the independence constraint, resulting in the larger contribution to the cost function of $1$ rather than $2 - M$.
In general, finding the optimal penalty $M$ is NP-hard~\cite{alessandroni2025alleviating}.
For the MIS problem this difficulty is largely circumvented, since the threshold above admits a simple closed form and a sufficiently strong penalty already enforces independence~\cite{ebadi2022quantum}. 
Taking $M$ arbitrarily large is nonetheless undesirable, because an excessively large penalty leads to loss of precision and numerical instability~\cite{cococcioni2021big}.
We therefore fix the penalty weight $M = 2$.
The cost Hamiltonian for MIS is obtained with the same procedure as for MC.
Due to their second-order objective functions, both MC and MIS are quadratic unconstrained binary optimization (QUBO) problems.

QAOA samples from the Ansatz
\begin{equation}\label{eq:Ansatz_state}
    \ket{\psi(\bm{\beta}, \bm{\gamma})} = \prod_{m=1}^p e^{-i \beta_m H_M} e^{-i \gamma_m H_C}\ket{+}^{\otimes n}
\end{equation}
with optimized variational parameters $(\bm{\beta}, \bm{\gamma})$ to find the ground state of $H_C$.
Here, we consider a problem-independent mixer operator \mbox{$H_M = -\sum_{i=0}^{n-1} X_i$}, although other mixing Hamiltonians have been proposed~\cite{fuchs2022constraint, he2023alignment}.

\subsection{\label{sec:swap_strategies}SWAP Strategies}

The transpilation task must determine an assignment of decision variables to physical qubits, known as the initial mapping, and find a routing procedure that enables all required interactions.
The inputs to this problem therefore include the target circuit and the hardware's connectivity, i.e. the coupling map $C$.
For $n$ qubits, $C$ is an $n \times n$ binary matrix whose entry $i,j$ is $1$ if qubit $i$ is physically coupled to qubit $j$ and $0$ otherwise.
For general circuits, standard transpilation approaches begin with a randomly generated initial qubit mapping, subsequently solve the routing problem, and then refine the initial mapping~\cite{li2019mapping, zou2024lightsabre}. 
For circuits composed of blocks of commuting gates, a structured transpilation strategy that exploits gate commutativity can reduce circuit depth.
The typical routing procedure for QAOA is thus a network of SWAP gates called a \emph{SWAP strategy} which is expressed through a sequence $\mathcal{S}=[S_{i_1}, ...,S_{i_K}]$ of $K=|\mathcal{S}|$ SWAP layers $S_{i_h}$ chosen from a set $\mathcal{B} = \{S_1, ..., S_{\ell}\}$ of $\ell$ SWAP layers~\cite{weidenfeller2202scaling}.
Here, $|\mathcal{S}|$ is the number of SWAP layers in $\mathcal{S}$.
A SWAP layer $S_i$ consists of SWAP gates that can be applied in parallel on the coupling map $C$. 
It induces a permutation $P_{i}$ on the physical qubits.
Each SWAP strategy is thus designed for a given hardware topology, since all SWAP layers are hardware-native. 
We say that a set of SWAP layers $\mathcal{B}$ forms a \emph{basis set} if there exists a sequence of SWAP layers built from $\mathcal{B}$ that enables arbitrary qubit interactions, see the top row in Fig.~\ref{fig:all_layers}.

\figureAllLayers

After $K$ SWAP layers and the initial mapping $\pi_0$, the implementable two-qubit connections $E_{0i_1i_2\ldots i_K}$ are determined by the logical conjunction of the couplings generated by the permuted coupling map
\begin{equation}\label{eq:implemented_edges}
    (\pi_0^\top C\pi_0) \bigvee_{k=1}^K (\pi_{i_k}^\top C\pi_{i_k}).
\end{equation}
Here, $\pi_{i_k} = P_{i_h} P_{i_{k-1}} \cdots P_{i_1} \pi_0$ is the qubit placement induced by the first $k$ SWAP layers of $\mathcal{S}$ and $\pi_0$, both of dimension $n \times 1$.
Thus, a SWAP strategy sequentially increases the set of implementable two-qubit interactions by creating different~$\pi_{i_k}$. 
In the following, we transpile QAOA circuits with different combinations of SWAP strategies $\mathcal{S}$ and coupling maps $C$ while keeping the rest of the transpilation pipeline identical. 
We thus refer to the transpilation methods we develop with two-tuples $(\mathcal{S}, C)$.

Given a particular SWAP strategy $\mathcal{S}$, the initial qubit mapping $\pi_0$ can be optimized to minimize the number of required SWAP layers.
The problem of deciding if there is an initial mapping that allows us to implement $H_C$ in $K$ SWAP layers of $\mathcal{S}$ can be mapped to a satisfiability (SAT) problem.
A binary search finds the minimal $K$ such that a valid mapping exists~\cite{matsuo2023sat}. 
We refer to this method as the \emph{SAT mapping}.
Although SAT problems are also NP-hard, there exist well-performing approximate classical solvers for them. 

Pre-determined SWAP layer sequences can reduce circuit depth for certain graph problems on hardware.
The most common strategy is the \emph{line} SWAP strategy $\mathcal{S}_\text{line}=[S_{1, \text{line}}, S_{2, \text{line}}, S_{1, \text{line}}, S_{2, \text{line}}, \ldots]$ applied to the linear coupling map $[C_\text{line}]_{i,j}=\delta_{i-1, j} \vee \delta_{i+1, j}$, where $i, j \in \{0, \dots, n-1\}$ index $n$ qubits and $\delta_{i,j}$ is the Kronecker delta.
We label this combination as $(\mathcal{S}_\text{line}, C_\text{line})$.
Here, the SWAP layers $S_{j, \text{line}}=\{(i, i+1)~|~ i\!\mod{2}=j\}$,
effectively alternately apply SWAP gates on even edges ($i\mod{2}=0$) and odd edges ($i\mod{2}=1$), see the top qubit rows of layers $S_1$ and $S_2$ in Fig.~\ref{fig:all_layers}, respectively. 
A fully connected QUBO with $n$ variables is optimally transpiled on a line using such $n-2$ alternating even and odd SWAP layers~\cite{weidenfeller2202scaling}.
Even on two-dimensional grid coupling maps $C_\text{grid}$, the line strategy is best for fully connected problems~\cite{weidenfeller2202scaling}.
Crucially, this strategy is not optimal for non-fully connected QUBOs. 
This limitation motivates the development of new SWAP strategies.

\section{SWAP strategies for grids\label{sec:swaps_for_grids}}

\subsection{Problem-independent strategies}\label{sec:problem_independent_2D_strategies}

Despite the additional connectivity of two-dimensional coupling maps, we often rely on the \emph{linear} transpilation strategy $(\mathcal{S}_\text{line}, C_\text{grid}^\text{line})$. 
Here, the effective coupling map $C_\text{grid}^\text{line}$ is the longest line of qubits in $C_{\mathrm{grid}}$, thereby reducing the problem to a one-dimensional structure. 

As a first contribution, we introduce a \emph{hybrid} transpilation strategy $(\mathcal{S}_{\mathrm{line}}, C_{\mathrm{grid}})$. It applies the line SWAP strategy on $C_\text{grid}^\text{line}$ while leveraging edges that are not on this line to apply $R_{ZZ}$ gates of $e^{-i\gamma H_C}$ when possible, see Fig.~\ref{fig:fig_hybrid}.
We optimize the initial mapping to further reduce the gate depth with the SAT mapping.
This hybrid strategy leverages the additional connectivity without modifying the underlying SWAP network.
It offers a useful intermediate baseline between line-based and two-dimensional SWAP strategies.

\figureLinearAndHybrid

We now consider the two-dimensional rectangular \emph{grid} transpilation strategy $(\mathcal{S}_{\mathrm{grid}}, C_{\text{grid}})$ of Ref.~\cite{weidenfeller2202scaling}, where the rectangular grid
$C_{\text{grid}}$ is defined by 
\begin{align}\label{eq:coupling_map_square}
    [C_\text{grid}]_{(i,j), (k,l)}=& \delta_{(i, j-1), (k, l)} \vee \delta_{(i, j+1), (k, l)} \\ & \vee \delta_{(i-1, j), (k, l)} \vee \delta_{(i+1, j), (k, l)}.\notag
\end{align}
Here, a qubit $(i, j)$ is indexed by its row $0\leq i<r$ and column $0\leq j<c$ in the grid.
The SWAP strategy $\mathcal{S}_{\mathrm{grid}}$ is constructed from a tailored basis set $\mathcal{B}_{\text{grid}}$ of SWAP layers, see top row in Fig.~\ref{fig:all_layers}. 
For a grid with $r$ rows and $c$ columns, $\mathcal{S}_{\mathrm{grid}}$ alternates horizontal SWAP layers $S_1$ and $S_2$ for $c-1$ steps, followed by vertical (row-swapping) layers $S_3$ and $S_4$. 
This sequence is then repeated $\lceil r / 2 \rceil$ times, yielding
\begin{equation}\label{eq:square_strategy}
    \mathcal{S}_{\mathrm{grid}} = \Big[\big((S_1, S_2)^{c-1}, S_3, S_4\big)^{\lceil r / 2 \rceil}\Big].
\end{equation}
Importantly, the alternating parity across rows in $S_1$ and $S_2$ allow qubits in neighboring rows to interact. 
In contrast to line-based approaches, this strategy fully exploits the grid connectivity.
Its performance---combined with the SAT mapping of Ref.~\cite{matsuo2023sat}---has not yet been studied. 

\subsection{\label{sec:protocol}Problem-dependent strategies}

We now introduce our main contribution, a \emph{greedy} problem-dependent transpilation to construct SWAP strategies and initial mappings tailored to the cost Hamiltonian.
We seek a sequence of SWAP layers $\mathcal{S}$, chosen from a basis $\mathcal{B}$, that has a smaller length $k$ than the length $k_{\mathrm{ref}}$ of a reference SWAP strategy $\mathcal{S}_\mathrm{ref}$.
We optimize $\mathcal{S}$ in an inner-loop and the initial mapping $\pi_0$ in an outer-loop.
The method takes as input the hardware coupling map $C_{\text{grid}}$ and a compatible basis set of SWAP layers $\mathcal{B}$.

First, an initial qubit mapping $\pi_0^{\mathrm{ref}}$ compatible with $\mathcal{S}_{\mathrm{ref}}$ is found following Ref.~\cite{matsuo2023sat, pysat-solvers-api}. 
Starting from $\pi_0^{\mathrm{ref}}$, see top node in Fig.~\ref{fig:algo}, the inner loop uses a depth-first-search (DFS) to explore the space of reachable qubit configurations.
These configurations form a tree whose branches correspond to possible SWAP layers from $\mathcal{B}$ and whose nodes correspond to qubit configurations $\pi_{0i_1 i_2 \ldots i_k}$, obtained after applying all the previous SWAP layers indexed by $i_1,i_2,\ldots,i_k$ to the initial mapping. 
Each node is also associated with a set of remaining target interactions $E^{\mathrm{rem}}_{0i_1 i_2 \ldots i_k}$, defined as the complement of the implemented interactions $E_{0i_1 i_2 \ldots i_k}$ from Eq.~\eqref{eq:implemented_edges} with respect to the full target edge set $E$. 
Since consecutive identical SWAP layers compose to the identity, all nodes have $|\mathcal{B}|-1$ children, except for the root node at $k=0$, which has $|\mathcal{B}|$ children; $|\mathcal{B}|=4$ in Fig.~\ref{fig:algo}.
The total number of nodes at depth $k$ is thus $|\mathcal{B}|\,(|\mathcal{B}|-1)^{k-1}$.

We seek the shortest path in this tree which implements all interactions, i.e., $E^{\mathrm{rem}}_{0i_1 i_2 \ldots i_k}=\varnothing$.
To avoid building this exponentially scaling tree, we do an exhaustive search only up to $k_{\max}$ SWAP layers at a time, see dashed line in Fig.~\ref{fig:algo}.
When the search reaches depth \(k_{\max}\) without implementing all required
interactions, the algorithm retains the partial SWAP-layer sequence that realizes the largest number of edges, with random tie-breaking, and continues from the resulting configuration. 
The depth-\(k_{\max}\) search is then repeated until all edges in \(E\) are implemented, after which the resulting SWAP sequence \(\mathcal{S}^j\) is returned, see bottom node of Fig.~\ref{fig:algo}. 
Index $j$ corresponds to the outer-loop iteration number.

If $k \geq k_\text{ref}$, the inner loop terminates and the outer loop computes a new initial mapping $\pi_0^j$ via the SAT mapping for the previous valid SWAP sequence, $\mathcal{S}^j = \mathcal{S}^{j-1}$. 
The greedy search in the inner loop then restarts from $\pi_0^j$. 
Because the SAT solver is stochastic, $\pi_0^j$ may differ from $\pi_0^{j-1}$, giving the method another chance to find a shorter SWAP strategy. 
Crucially, each iteration cannot increase the circuit depth, since a solution of at most the previous depth is already known.

For grid coupling maps, we set $\mathcal{S}_{\mathrm{ref}} = \mathcal{S}_{\mathrm{grid}}$, and $\mathcal{B} = \mathcal{B}_{\mathrm{grid}} (\text{or } \mathcal{B}_{\text{grid}}^{\text{extended}} = \mathcal{B}_{\mathrm{grid}} \cup \{ S_5, S_6, S_7, S_8\})$, see Fig.~\ref{fig:all_layers}.
Additional SWAP layers can be included to trade search overhead for reduced circuit depth.
The detailed algorithmic steps and impact of the hyperparameter $k_\mathrm{max}$ are in App.~\ref{sec:hyperparameters_App.}.

\figureAlgo

\section{\label{sec:results}Results}

\figureCircuitMetrics
\figureHeatmaps

\subsection{\label{sec:simulations}Simulation Benchmarks}

We now benchmark the SWAP strategies introduced in Sec.~\ref{sec:swap_strategies} and \ref{sec:swaps_for_grids} on two-dimensional rectangular grid coupling maps $C_{\text{grid}}$.
For a given optimization problem, we chose lattice dimensions as close to square as possible, potentially leaving some qubits unused. 
Details on the impact of the lattice aspect-ratio are in App.~\ref{sec:squareness}.
We measure the \emph{complexity} of a circuit implementing $e^{-i\gamma H_C}$ by (i) the number of SWAP layers, (ii) the two-qubit gate depth, and (iii) the number of two-qubit entangling gates. 
All metrics thus correspond to depth-one QAOA. 

The \text{linear} strategy uses a subset of edges that form a line $C_{\text{line}}$.
The greedy strategy uses the extended SWAP layer basis $\mathcal{B}_{\text{grid}}^{\text{extended}}$, with fixed parameters $k_{\max}=5$ and $I=5$ iterations of the outer loop. 
We benchmark with Random Regular (RR) graphs with degree $d=3$, also called $d-$regular graphs, and Erdős–Rényi (ER) graphs with edge probability $q=0.08$ with 40 to 90 nodes.
The density of RR graphs is inversely proportional to the number of nodes. 
The density of Erdős–Rényi graphs is on average $q$, i.e. constant. 
We consider these three-regular and 8\% edge probability graphs to be sparse.

On $C_\text{line}$, the \emph{linear} strategy achieves substantially lower circuit depths than the SABRE algorithm implemented in Qiskit~\cite{li2019mapping, zou2024lightsabre}. 
Here, the depth of SABRE for 40 to 90 nodes ranges from 149 to 497 for RR graphs and from 236 to 1530 for ER graphs, data not shown in Fig.~\ref{fig:circuit_metrics}(b) and (e). 
By contrast, the two methods yield comparable CZ-gate counts, see Fig.~\ref{fig:circuit_metrics}(c) and (f). 
When a grid coupling map is available, all grid-based strategies produce improved circuits relative to their corresponding line version, compare the \emph{hybrid}, \emph{grid}, and \emph{greedy} curves with the \emph{linear} curve in Fig.~\ref{fig:circuit_metrics}, and the brown and dark blue curves in Fig.~\ref{fig:circuit_metrics}(c) and (f). 
Notably, even on a grid coupling map, SABRE produces circuits with substantially greater depth than all specialized grid-based SWAP strategies, see Fig.~\ref{fig:circuit_metrics}(b) and (e), but reduces CZ gate count, see Fig.~\ref{fig:circuit_metrics}(c) and (f). 
We therefore use the \emph{linear} strategy as the principal benchmark throughout the remainder of the analysis.

The \emph{linear} strategy requires the largest number of SWAP layers and CZ gates, see Fig.~\ref{fig:circuit_metrics}.
The \emph{hybrid} strategy is better than the \emph{linear} one: all three metrics are lower by constant factors.
This improvement is smaller for ER graphs than three-regular ones, compare Fig.~\ref{fig:circuit_metrics}(a, b, c) to (d, e, f), respectively.
The number of SWAP layers, depth, and CZ gates scale as $n$, $n$, and $n^2$, respectively, for the three-regular graphs with both strategies.
ER graphs require an additional factor of $n$.
The complexity metrics of the \emph{grid} and \emph{greedy} strategies are lower than the line-based ones.
Due to its problem-adaptive search, the \emph{greedy} strategy further reduces the number of SWAP layers, depth, and CZ gate count over the \emph{grid} strategy.
This improvement is more pronounced when the graphs have more than $n=55$ nodes, see Fig.~\ref{fig:circuit_metrics}.
Additional results for four-regular graphs and ER graphs with 6.5\% edge probability support the same trends, see App.~\ref{sec:other_parameters_results}.
There, we also compare these SWAP strategies to the out-of-the-box Qiskit transpiler on individual three-regular instances with $100$ nodes.
 
Crucially, the resources needed are determined by whether the SWAP network routes qubits along a line or the two-dimensional grid, see Fig.~\ref{fig:circuit_metrics}. 
The \emph{linear} and \emph{hybrid} strategies therefore exhibit similar behavior in the graph size $n$. 
Importantly, the \emph{grid} and \emph{greedy} strategies achieve an asymptotic $O(\sqrt{n})$ improvement in all relevant circuit-complexity metrics, consistent with the analysis in App.~\ref{sec:asymptotic_analysis}. 
However, the $O(\sqrt{n})$ advantage of grid-based coupling maps is specific to sparse graphs. 
For sufficiently dense graphs, the \emph{linear} strategy is optimal~\cite{weidenfeller2202scaling}. 
We estimate empirically the graph sizes and densities for which the proposed grid-based methods are expected to be advantageous in App.~\ref{sec:asymptotic_analysis}.

We now quantify the advantage of the \emph{greedy} method over the \emph{linear} one by transpiling $e^{-i\gamma H_C}$ for ER graphs with $n \in \{50,60,\ldots,100\}$ and edge probabilities $q \in \{0.04,0.06,\ldots,0.14\}$. 
We consider circuits with fewer than $3{,}600$ CZ gates to have a tolerable noise level, see App.~\ref{sec:errors}. We identify such circuits using the dashed lines in Fig.~\ref{fig:heatmaps}(a) and (b). 
The \emph{greedy} approach makes instances within reach of the \emph{linear} strategy more noise robust by yielding shallower circuits. 
Crucially, it also extends the size and density of instances we identify as feasible, see the region between the dashed lines in Fig.~\ref{fig:heatmaps}(c).

As hardware improves, both feasibility boundaries will shift toward larger problem sizes. 
For sparse ER graphs, the advantage of the \emph{greedy} strategy over the \emph{linear} one increases because its CZ count grows as $O(n^{5/2})$ with problem size instead of $O(n^3)$. 
Thus, a fixed increase in the available CZ budget translates into a larger increase in accessible problem size for the \emph{greedy} strategy.
For example, at edge density $q=0.08$, increasing the feasible budget to $5{,}500$ CZ gates extends the accessible problem size from approximately 60 to 70 nodes with the \emph{linear} strategy, see Fig.~\ref{fig:heatmaps}(a), but from about 80 to 100 nodes with the \emph{greedy} strategy, see Fig.~\ref{fig:heatmaps}(b). 
This illustrates how hardware improvements amplify the benefits of efficient transpilers.
 
\figureTruncatedImplementations
\figureHardwareResultsAll

Our \emph{greedy} strategy also outperforms existing approaches which truncate the cost Hamiltonian in the QAOA Ansatz, see Fig.~\ref{fig:truncated}. 
Truncated Ansatz sacrifice the theoretical performance guarantee of QAOA in the $p\to\infty$ limit to reduce circuit depth~\cite{druagoi2026approximate}.
This yields better overall performance on noisy hardware. 
Previous work shows that this trade-off is beneficial for three-regular graphs when the truncation is tailored to the hardware.
For instance, by restricting the interactions to those reachable within a limited number of \emph{linear} SWAP layers~\cite{druagoi2026approximate}.
The \emph{greedy} strategy implements a significantly larger subset of the target interactions for the same number of SWAP layers than the linear and hybrid methods. 
For example, for three-regular graphs and $15$ allowed SWAP layers, the \emph{greedy} approach captures all required interactions, whereas the \emph{linear} strategy implements only about half, see Fig.~\ref{fig:truncated}(a). 
A similar behavior also holds for the ER graphs with $q=0.08$. 
With $20$ SWAP layers, the \emph{greedy} approach captures about $90\%$ of all required interactions, whereas the \emph{linear} strategy implements only about half, see Fig.~\ref{fig:truncated}(b). 
Consequently, given a budget circuit depth, our method retains more problem structure than the \emph{linear} strategy.

\subsection{\label{sec:hardware}Hardware Results}

We now experimentally probe the extended feasibility regime by executing QAOA circuits for the MC and MIS problems on IBM's rectangular-lattice hardware for graphs with up to $80$-nodes. 
Performance is evaluated using the mean approximation ratio. 
For a given bitstring $x$, we define the approximation ratio as
\begin{equation}
    r(x) = \frac{f(x)}{\max_{x'} f(x')} \in [0,1],
\end{equation}
where $\max_{x'} f(x')$ is the optimal classical cost, and $f(x)$ is either $f_{\text{MC}}$ or $f_{\text{MIS}}$, see Eqs.~\eqref{eq:cost_function_MC} and~\eqref{eq:cost_function_MIS}.
Optimal samples $x^\star$ satisfy $r(x^\star)=1$.
The quality of $s$ samples $\{x_i\}_{i=1}^s$ is the mean approximation ratio
\begin{equation}
    \bar{r} = \frac{1}{s}\sum_{i=1}^s r(x_i).
\end{equation}

First, we consider the MC problem on RR graphs with degree $d=3$. 
For such instances, uniformly sampled candidate solutions yield  
$\bar{r}\approx 0.54$, see App.~\ref{sec:uniform_perf} and black line in Fig.~\ref{fig:hardware_results_all}(a).
The best known polynomial-time approximation algorithm, the Goemans-Williamson algorithm~\cite{goemans1995improved}, produces samples with at least $\bar{r} \gtrsim 0.878$.
For QAOA, rigorous performance guarantees on three-regular graphs yield approximation ratios of $\bar{r} \geq 0.692$ for $p=1$ and $\bar{r} \geq 0.756$ for $p=2$ layers~\cite{wurtz2021fixed, basso2021quantum, MC_lower_bounds}, see dashed horizontal lines in Fig.~\ref{fig:hardware_results_all}(a).

The QAOA angles in our circuits are the $p=1$ and $p=2$ fixed angles of Ref.~\cite{wurtz2021fixed}  used without further optimization. 
We also compare the hardware results to samples drawn from the matrix product state (MPS) simulator of Qiskit Aer~\cite{javadi2024quantum} with a bond dimension of 64.
As $n$ increases, the mean approximation ratio of the MPS results decrease and eventually fall under the theoretical lower bound, see Fig.\ref{fig:hardware_results_all}(a).
We attribute this to approximations in the MPS.

Hardware noise deteriorates the quality of the quantum samples. 
Crucially, our \emph{greedy} method consistently outperforms the standard \emph{linear} SWAP strategy across all tested system sizes as measured by the average approximation ratio, compare dashed and dotted lines in Fig.~\ref{fig:hardware_results_all}(a). 
These results are obtained without post-selection, which is discussed and tested in App.~\ref{sec:experimental_postprocessing}, or error suppression~\cite{viola1999dynamical, biercuk2009optimized, niu2022effects, pokharel2018demonstration}.
The cumulative distribution functions $\mathrm{CDF}(c) = \mathrm{Prob}(f(x) \leq c) \in [0,1]$ of the sampled cost values $f(x)$ further illustrates this gain, see Fig.~\ref{fig:hardware_results_all}(b). 
Since higher cost values are desirable, a right-shifted CDF indicates an improved performance.
Noiseless MPS simulations show a rightward shift when increasing the QAOA depth from $p=1$ to $2$. 
However, hardware noise reverses this trend for $80$-node graphs, where $p=1$ outperforms $p=2$. 
Nevertheless, our \emph{greedy} method consistently yields distributions with higher cost values than the \emph{linear} strategy, compare dashed and dotted lines in Fig.~\ref{fig:hardware_results_all}(b).

As $n$ increases, the mean approximation ratios of the circuits transpiled with the \emph{greedy} strategy do not deteriorate as quickly as those obtained with the \emph{linear} strategy on hardware.
We quantify this effect through the relative increase in approximation ratio,
\begin{equation}
    \delta \bar r = \frac{\bar r^\text{greedy} - \bar r^\text{linear}}{\bar r^\text{linear}} = \frac{\Delta \bar r}{\bar r^\text{linear}},
\end{equation}
which equals the relative increase in cut size $\Delta c / c^\text{linear}$, since the approximation ratio is the cut value normalized by the optimal cut, see Fig.~\ref{fig:hardware_results_all}(c). 
The $\delta \bar r$ reaches up to $6.62\%$ for $n = 72$ nodes. 
At this size, we measure an average $\bar r^\text{linear} = 0.60$, corresponding to a cut value of $c^\text{linear} =  \bar r^\text{linear} \cdot c^\text{optimal} \approx 59$. 
The $6.62\%$ improvement thus implies that
$\Delta c = \delta \bar r \cdot c^\text{linear} \approx 4$
more edges are cut, on average.
The relative improvement of \emph{greedy} over \emph{linear} is typically larger for $p=2$ than for $p=1$ since it avoids nearly twice as many gates at $p=2$ than at $p=1$, see Fig.~\ref{fig:hardware_results_all}(d).

We now analyze the impact of an efficient transpilation on MIS problems on ER graphs with an $8\%$ edge probability. 
These problems are harder than MC on three-regular graphs, due to (i) the increased circuit complexity, (ii) the constraints of MIS, and (iii) the harder QAOA angle optimization. 
First, their expected edge density is 8\%, i.e., more than double the $3.8\%$ of the 80-node three-regular graphs.
This increased density and the arbitrary edge structure make the QAOA circuits more complex.
Second, the independent set constraint requires a penalty, see Sec.~\ref{sec:qaoa}, and also implies that many solutions are infeasible.
We post-process infeasible samples to enforce independence~\cite{MIS_postprocessing}, and then evaluate approximation ratios only on the set of viable solutions. 
Finally, the QAOA parameter optimization is more challenging for MIS than MC on RR graphs, see App.~\ref{sec:param_optimization}.

Crucially, despite these additional challenges, our \emph{greedy} transpilation method yields consistently better approximation ratios than the \emph{linear} one across all tested problem sizes and QAOA depths, see Fig.~\ref{fig:hardware_results_all}(e). In addition, our method also yields consistently better sampling distributions, see Fig.~\ref{fig:hardware_results_all}(f). 
We observe relative improvements of up to $\delta \bar r = 9.32\%$ for $56$-node instances at $p=2$, where greedy transpilation reduces the circuit to an average number of $2412$ CZ gates (standard deviation $472$) —roughly half the $4612$ (standard deviation $267$) required by the linear method, see Fig.~\ref{fig:hardware_results_all}(g). 
At this size, we measured an average 
independent set of size $c^\text{linear} \approx 16$. The improvement in the size of the independent set is $\Delta c = \delta \bar r \cdot c^\text{linear} \approx 1.5$ nodes. 
Notably, this corresponds to the largest MIS circuit tested, measured by gate-count, and thus the largest absolute gate reduction, see Fig.~\ref{fig:hardware_results_all}(h). 
The increasing relative improvement with circuit complexity shows the benefits of a problem-aware transpilation for larger and deeper circuits, see Fig.~\ref{fig:hardware_results_all}(h). 

\section{\label{sec:conclusion}Conclusions and Discussion}

We study transpilation methods on two-dimensional lattices for circuits with sparse blocks of commuting two-qubit gates.
Such circuits commonly arise in quantum algorithms for combinatorial optimization and simulating Ising models. 
First, we extend the standard \emph{linear} routing SWAP strategy to a \emph{hybrid} one by applying interactions on the grid edges that are not part of the line. 
This reduces circuit complexity without changing its scaling with size.
This motivates designing SWAP strategies for grid coupling maps.
Second, we combine the grid strategy of Ref.~\cite{weidenfeller2202scaling} with the initial mapping of Ref.~\cite{matsuo2023sat}.
This further reduces circuit complexity and crucially results in a more favorable scaling with system size for sparse graphs.
Third, we develop a \emph{greedy} strategy to tailor the ordering of SWAP layers to the problem instance. 
This further reduces circuit complexity by minimizing the number of SWAP layers, and consequently also circuit depth and the number of entangling gates. 

When specialized to rectangular lattices in our numerical experiments, the \emph{greedy} method yields substantial reductions in circuit complexity for instances with up to $80$ qubits, at a tunable classical preprocessing cost worth incurring to reduce the
two-qubit gate count on noisy hardware.
The amount of additional classical pre-processing is controlled by restricting the search depth, see App.~\ref{sec:hyperparameters_App.}.
Importantly, the reduction in the number of CZ gates increases with QAOA depth and graph complexity. 
These improvements extend the range of problem sizes that noisy hardware can meaningfully execute. 
For ER graphs with densities $q \in [0.04, 0.14]$, we estimate an increase of approximately 20 variables in problem size.
Hardware experiments corroborate these findings, showing performance gains of up to $6.62\%$ in the approximation ratio for MC and up to $9.32\%$ for MIS.

Hardware improvements are expected to be especially pronounced for problem-aware transpilation on problems sparse enough to leverage the $\sqrt{n}$ graph diameter of grid coupling maps. 
Thus, improvements in hardware quality should not only increase the absolute size of tractable instances, but also amplify the practical advantage of routing strategies that exploit the two-dimensional connectivity.
This makes our approach particularly relevant for testing quantum algorithm development and probing regimes of potential quantum advantage, which include hardness regimes deemed interesting for exploring quantum advantage. 
In the QOBLIB benchmark of Ref.~\cite{koch2025decathlon}, a state-of-the-art classical solver fails to certify several MIS instances as optimal within the benchmark time limit~\cite{hard_instances}, the smallest low-density case having $500$ nodes and an edge density of about $5\%$. 

Since our method relies on predefined SWAP layers, future work could design improved SWAP layer basis sets. 
For instance, by combining adjacent SWAP layers that are not separated by parametrized gates, or by leveraging AI-based tools.
These alternative choices may further reduce circuit complexity.
In addition, our approach could be extended to alternative hardware topologies. 
Recent advances in quantum error correcting codes, such as the bivariate bicycle code, rely on hardware connectivity based on Tanner graphs~\cite{bravyi2024high, yoder2025tour, eberhardt2024logical, eberhardt2024pruning}. These architectures require rectangular lattices augmented with additional long-range connections. 
Adapting transpilation strategies to such structured, non-local connectivity may also reduce circuit complexity. 

Overall, our results highlight the importance of problem-aware transpilation and suggest that such strategies are key to improve the performance of quantum algorithms on noisy devices.

\section*{Acknowledgments}

The authors thank Stefan Woerner and Anthony Gandon for helpful discussions, and Ehud Karavani for valuable comments on the manuscript. 
S.D. thanks Laurin Fischer for software support related to hardware experiments. 
This work made use of IBM Quantum services.
S.D. acknowledges funding from the ETH Zurich Quantum Center.

\appendix 
\numberwithin{equation}{section}

\section{Details of the greedy algorithm and hyperparameter tuning}\label{sec:hyperparameters_App.}

\setcounter{figure}{0}
\renewcommand{\thefigure}{A\arabic{figure}}

Here, we give the steps of our \emph{greedy} transpilation method and analyze its classical cost. 
We also introduce a variant that can improve performance at a smaller runtime cost than increasing the search depth.

\subsection{Algorithms}\label{subsec:algorithms_App.}

The outer loop, Alg.~\ref{alg:protocol}, finds the initial mapping for a given SWAP strategy $\mathcal{S}^j$.
For the initial mapping $\pi_0^{j-1}$ at iteration $j$, the inner loop, Alg.~\ref{alg:greedy_search}, constructs the SWAP strategy $\mathcal{S}^j$ by exploring the tree of reachable qubit configurations.

In the standard greedy method, the full selected path is committed before the search resumes from its endpoint, i.e. $k_{\text{append}} = k_{\max}$.
We introduce a variant that instead commits only $k_{\text{append}} \leq k_{\max}$ SWAP layers and resumes the search from the corresponding intermediate node rather than the final leaf, see the blue lines in Alg.~\ref{alg:greedy_search}.
This partial commitment reduces the rigidity of the greedy choice.
Although paths are still selected on short-term gain, committing a shorter prefix allows adjustments in subsequent steps and can improve performance.
Reaching such gains through a smaller $k_{\text{append}}$ is typically cheaper than reaching comparable gains by increasing the lookahead depth $k_{\max}$, see App.~\ref{subsec:cost_App.}.

\begin{algorithm}[H]
\caption{Problem-dependent circuit transpilation}
\label{alg:protocol}
\begin{algorithmic}[1]
\Require 
\Statex target interaction graph $G=(V,E)$
\Statex hardware coupling map $C_{\text{grid}}$
\Statex basis set of SWAP layers $\mathcal{B}$
\Statex maximum greedy exploration depth $k_{\max}$
\Statex append depth $k_{\text{append}} \leq k_{\max}$
\Statex number of refinement iterations $I$
\Ensure ($\mathcal{S}_{\mathrm{greedy}}$, $\pi_0^{\mathrm{greedy}}$)

\State Define a fixed reference SWAP sequence $\mathcal{S}_{\mathrm{ref}}$ that can implement any pairwise interaction.
\State Use a SAT-solver to find an initial qubit mapping $\pi_0^{\mathrm{ref}}$.
\State Compute the depth bound $k_{\mathrm{ref}} \gets |\mathcal{S}_{\mathrm{ref}}|$.
\State $\mathcal{S}_{\mathrm{greedy}} = \mathcal{S}^0 \gets \mathcal{S}_{\mathrm{ref}}$, \quad $\pi_0^{\mathrm{greedy}} = \pi_0^0 \gets \pi_0^{\mathrm{ref}}$

\For{$j=1$ to $I$}
    \State $\mathcal{S}^j \gets \textsc{GreedySearch}(G, C, \mathcal{B}, \pi_0^{j-1}, k_{\max}, k_{\text{append}}, k_{\mathrm{ref}})$
    \If{$\mathcal{S}^j$ == FAIL}
        $\mathcal{S}^j = \mathcal{S}^{j-1}$
    \EndIf
    \State Use a SAT-solver to compute a new $\pi_0^j$ for $\mathcal{S}^j$.
    \If{$|\mathcal{S}^j| < |\mathcal{S}_{\mathrm{greedy}}|$}
        \State $\mathcal{S}_{\mathrm{greedy}} \gets \mathcal{S}^j$
        \State $\pi_0^{\mathrm{greedy}} \gets \pi_0^j$
        \State $k_{\mathrm{ref}} \gets |\mathcal{S}^j|$
    \EndIf
\EndFor

\State \Return $(\mathcal{S}_{\mathrm{greedy}}, \pi_0^{\mathrm{greedy}})$
\end{algorithmic}
\end{algorithm}

\begin{algorithm}[H]
\caption{\protect\textsc{GreedySearch} over SWAP strategies}
\label{alg:greedy_search}
\begin{algorithmic}[1]
\Require 
\Statex target interaction graph $G=(V,E)$
\Statex hardware coupling map $C_{\text{grid}}$
\Statex basis set of SWAP layers $\mathcal{B}$
\Statex initial qubit mapping $\pi_0^{j-1}$
\Statex maximum greedy exploration depth $k_{\max}$
\Statex \textcolor{ibmblue}{append depth $k_{\text{append}} \leq k_{\max}$}
\Statex total depth budget $k_{\mathrm{ref}}$
\Ensure SWAP sequence $\mathcal{S}^j$

\State $\mathcal{S}^j \gets [\,]$, \quad $\pi \gets \pi_0^{j-1}$, \quad $E^{\mathrm{rem}} \gets E$

\While{$E^{\mathrm{rem}} \neq \emptyset$}
    \State Construct a tree rooted at $(\pi, E^{\mathrm{rem}})$ by exploring all SWAP-layer sequences over $\mathcal{B}$ up to depth $k_{\max}$.
    \If{there exists a leaf with no remaining interactions}
        \State Choose the shallowest such leaf.
        \State Append the corresponding SWAP path to $\mathcal{S}^j$.
        \State \Return $\mathcal{S}^j$
    \Else
        \State Randomly choose one of the leaves that implements the largest number of remaining interactions.
        \State \textcolor{ibmblue}{Append $k_{\text{append}}$ SWAP layers from chosen path.} 
        \State Update $(\pi, E^{\mathrm{rem}})$ to the node reached \textcolor{ibmblue}{after $k_{\text{append}}$}.
    \EndIf
    \If{$|\mathcal{S}^j| > k_{\mathrm{ref}}$} 
        \Return FAIL 
    \EndIf
\EndWhile

\State \Return $\mathcal{S}^j$
\end{algorithmic}
\end{algorithm}

\figureHyperparameters

\subsection{Computational cost and hyperparameter tuning}\label{subsec:cost_App.}

We now investigate the trade-off between performance and computational cost of the \emph{greedy} method.
When $k_{\text{append}} = k_{\max}$, it explores a number of paths in the tree that scales as
\begin{equation}
    T_{\textrm{standard}} \sim I \cdot N_\text{subtrees} \cdot |\mathcal{B}| \cdot (|\mathcal{B}|-1)^{k-1}.
    \label{eq:greedy_scaling}
\end{equation}
Here, $I$ is the number of outer-loop iterations and $N_{\text{subtrees}}$ is the number of subtrees explored until a sequence of SWAP layers that implements all target interactions is found.
Although $N_{\text{subtrees}}$ is not known a priori, it is expected to grow with the number of target interactions, since more interactions typically require longer SWAP sequences and thus more subtrees under the depth constraint $k_{\max}$.
For small $k_{\max}$, e.g. $\leq 4$, the limited lookahead performs poorly.
Increasing $k_{\max}$ enables a deeper lookahead and thus reduces the number of SWAP layers in the solution, see the diagonals in Fig.~\ref{fig:hyperparams}(a,c).
However, a large $k_{\max}$ is computationally prohibitive, see Eq.~\eqref{eq:greedy_scaling} and the diagonals in Fig.~\ref{fig:hyperparams}(b,d).
These data suggest that good performance is achievable with a moderate $k_{\max}$.

Setting $k_{\text{append}}<k_{\max}$ can improve performance, but requires exploring a larger number of subtrees $N'_\text{subtrees}$. 
This number is roughly determined by the total number of qubit permutations of the final SWAP sequence 
\begin{equation}
    k \cdot N_\text{subtrees} \approx k_\text{append} \cdot N'_\text{subtrees}.
    \label{eq:nsubtrees_relation}
\end{equation}
By combining Eqs.~\eqref{eq:nsubtrees_relation} and~\eqref{eq:greedy_scaling} we see that the number of paths for $k_{\text{append}} < k_{\max}$ scales as 
\begin{equation}
    T_{\textrm{variant}} \sim I \cdot \frac{k}{k_\text{append}} \cdot N_\text{subtrees} \cdot |\mathcal{B}| \cdot (|\mathcal{B}|-1)^{k-1}.
    \label{eq:cost_adapt}
\end{equation}
Despite the prefactor $k/k_\text{append} \geq 1$, selecting $k_\text{append}<k$ can be cheaper than directly increasing the lookahead depth $k$ by $\Delta k$ which incurs an exponential cost $(|\mathcal{B}|-1)^{\Delta k}$.
Therefore, reducing $k_\text{append}$ at fixed $k$ is preferable from a runtime perspective when
\begin{equation}
    \frac{k}{k_\text{append}} \leq (|\mathcal{B}|-1)^{\Delta k}.
    \label{eq:tradeoff_condition}
\end{equation}
For example, when deciding between increasing the subtree search depth by $\Delta k = 1$ or reducing the appended path length by one, i.e., $k_\text{append} = k - 1$, Eq.~\eqref{eq:tradeoff_condition} becomes $k/(k-1) \leq |\mathcal{B}|-1$ which holds for any basis with $|\mathcal{B}| \geq 3$.
This trend is supported by Fig.~\ref{fig:hyperparams}(b,d).

\section{Impact of lattice aspect-ratio}\label{sec:squareness}
\setcounter{figure}{0}
\renewcommand{\thefigure}{B\arabic{figure}}

\figureSquarenessImpactAppendix

Here, we model the circuit complexity of a QAOA cost layer for a QUBO corresponding to a graph $G = (V,E)$ with $n=|V|$ nodes, transpiled onto a rectangular coupling map. Section~\ref{subsec:general_setup} develops a general estimate of the circuit depth and gate count.
Sec.~\ref{sec:aspect_ratio} then specializes to a rectangular lattice of fixed size and varying aspect-ratio, and shows that the depth is smallest for near-square layouts. 

\subsection{General Set-Up}
\label{subsec:general_setup}

For a given placement of decision variables to qubits, called a \emph{qubit permutation}, only a subset of the target $R_{ZZ}$ gates can be executed without SWAPs in what we call an \emph{interaction block}.
Since a qubit cannot simultaneously participate in multiple gates, the two-qubit gate depth of such a block $D_{\mathrm{int}}$ is upper bounded by the maximum degree of the hardware coupling map $\bar{d}$.
To implement the remaining target interactions, the decision variables must be routed into a new permutation, and the process repeated until all target gates are executed. 
The final circuit thus consists of $N_{\text{perm}}$ interaction blocks and \emph{routing blocks} which permute the qubits. 
If each routing block has the same depth $D_{\mathrm{route}}$, the circuit depth is
\begin{equation}\label{eq:circuit_depth} 
D = N_{\text{perm}} \times \bigl(D_{\mathrm{int}}+D_{\mathrm{route}}\bigr).
\end{equation}
We upper bound the total number of entangling gates $N$ in the final transpiled circuit by $O(nD)$ since $D$ dense layers of two-qubit gates contains $nD/2$ gates. 
Similarly, the total gate count of routing and interaction blocks is bounded by $O(nD_{\textrm{route}})$ and $O(nD_{\textrm{int}})$, respectively.

\subsection{Aspect-ratio impact}
\label{sec:aspect_ratio}

We study the aspect-ratio dependence of a circuit's depth and gate-count for sparse graphs with $|E| = \Theta(n)$ edges.
We consider a rectangular coupling map with $n_r$ rows and $n_c =  \lceil n/n_r \rceil \geq n_r$ columns, leaving only a few unused qubits. 
The average node degree of the rectangular lattice with nodes $\mathcal{V}$ and connection edges $\mathcal{E}$,
\[
\bar{d} = \frac{2|\mathcal{E}|}{|\mathcal{V}|} = \frac{2\big(n_r(n_c-1) + n_c(n_r-1)\big)}{n_r n_c} = 4 - \frac{2}{n_r} - \frac{2}{n_c},
\]
starts at $2 - 2 /n$ for the line $(n_r,n_c)=(1,n)$ to  approach  $4 - 4/\sqrt{n}$ for the square $n_r = n_c = \sqrt{n}$. 

The maximum number of edges from $E$ that an interaction block can implement is $n\bar{d}/2$, i.e., each of the $\bar d$ interaction layers contains $n/2$ simultaneous $R_{ZZ}$ gates.
Therefore, $N_{\mathrm{perm}}$ is bounded from below by $2|E|/(n\bar{d})$.

For a fixed problem instance, the interaction blocks must implement all the target gates.
The leading aspect-ratio dependence is thus expected to come from the routing block.
For instances with $|E|=\Theta(n)$, target interactions are rare enough that routing blocks bring a few specific pairs of qubits into contact. 
The depth $D_{\mathrm{route}}$ should thus scale with the diameter of the coupling map $\mathscr{D}(C)=n_r+n_c$, i.e., the longest shortest path between any two nodes of $C$. 
We join these considerations to model the circuit depth, i.e., Eq.~\eqref{eq:circuit_depth}, of graphs with $|E| = \Theta(n)$ edges 
by the function 
\begin{equation}\label{eq:layer_depth}
    D(n_r) = a_D + b_D\Big({n_r} + \frac{n}{n_r}\Big) \Big/ \Big(4 - \frac{2}{n_r} - \frac{2n_r}{n} \Big).
\end{equation}
The coefficients $a_D, b_D$ minimize the mean squared error with respect to simulated data.
The same functional form fits the entangling-gate count, since $N = O(nD)$ and the factor $n$ is independent of the aspect-ratio, i.e.
\begin{equation}\label{eq:layer_volume}
    N(n_r) = a_N + b_N \Big({n_r} + \frac{n}{n_r}\Big) \Big/ \Big(4 - \frac{2}{n_r} - \frac{2n_r}{n} \Big).
\end{equation}

We route RR graphs with degrees three and four, and ER graphs with $6.5\%$ and $8\%$ edge densities. 
All graphs have $n=100$ nodes. 
We consider lattices with $n_r\in\{1, ..., 10\}$ and $n_r\times n_c\geq100$.
The circuit metrics follow the proposed scaling across all graph families, see Fig.~\ref{fig:squareness_impact}, with large coefficients of determination~$R$.
Once normalized by the number of edges, the per-edge offset $a/|E|$ agrees across graph families to within about $10\%$ for both the depth and the gate count, see Tab.~\ref{tab:squareness_fit_parameters_edges}.
This confirms that the interaction cost is set by the problem rather than the lattice. 
Crucially, the line is strongly penalized by the routing factor $(n_r + n_c)/\bar{d}$. 
Therefore, a genuinely two-dimensional layout is preferable to a line with square aspect ratio being optimal for sparse graphs.

\begin{table}[t]
\centering
\setlength{\tabcolsep}{3pt}
\begin{ruledtabular}
\begin{tabular}{lcccccc}
\multirow{2}{*}{\textbf{Graph}} & \multicolumn{2}{c}{\textbf{\# SWAP layers}} & \multicolumn{2}{c}{\textbf{Depth}} & \multicolumn{2}{c}{\textbf{\# CZ gates}} \\
\cline{2-7}
 & $a$ & $a/|E|$ & $a\,(10^3)$ & $a/|E|$ & $a\,(10^3)$ & $a/|E|$ \\
\hline
RR $d=3$        & 8.65  & 0.06 & 0.05 & 0.32 & 1.36 & 9.05  \\
RR $d=4$      & 14.43 & 0.07 & 0.08 & 0.38 & 2.12 & 10.59 \\
ER $0.065$ & 19.39 & 0.06 & 0.12 & 0.35 & 2.83 & 8.80  \\
ER $0.08$  & 34.12 & 0.09 & 0.15 & 0.38 & 4.03 & 10.18 \\
\hline
Mean & -- & 0.07 & -- & 0.36 & -- & 9.65 \\
SD   & -- & 0.01 & -- & 0.03 & -- & 0.87 \\
\end{tabular}
\end{ruledtabular}
\caption{\justifying Fit parameters $a$ and per-edge offset $a/|E|$ from the aspect-ratio fits in Fig.~\ref{fig:squareness_impact}, using Eq.~\eqref{eq:layer_depth} for SWAP layers and depth and Eq.~\eqref{eq:layer_volume} for CZ gates. Summary rows give the mean, standard deviation (SD).
}
\label{tab:squareness_fit_parameters_edges}
\end{table}

\section{\label{sec:asymptotic_analysis}Asymptotic Advantage of Transpiling Sparse Graphs on Rectangular Lattices}
\setcounter{figure}{0}
\renewcommand{\thefigure}{C\arabic{figure}}

We call an interaction graph sparse when its size and density place it in the regime where the grid-like connectivity is advantageous over a line, and dense otherwise.
We use the routing model of Sec.~\ref{subsec:general_setup} to justify in Sec.~\ref{subsec:sparse_regime} the resource scalings for sparse instances seen in Fig.~\ref{fig:circuit_metrics}. 
Dense instances are discussed in Sec.~\ref{subsec:dense_regime}. 
We reconcile both regimes in Sec.~\ref{subsec:crossover}. 

\figureConceptualGraphRegimes

\subsection{Sparse regime}
\label{subsec:sparse_regime}

We call the target graph sparse when the routing contribution in Eq.~\eqref{eq:circuit_depth} dominates the circuit depth. 
This occurs when the number of required qubit permutations, $N_{\mathrm{perm}}$, remains small compared with the routing depth which depends on $\mathscr{D}(C)$, see Fig.~\ref{fig:conceptual_graph_regimes}(a). 
$d$-regular graphs $G$ with $d\ll \mathscr{D}(C)$ satisfy $N_{\mathrm{perm}}\ll\mathscr{D}(C)$. 
Indeed, RZZ gates can be applied in parallel if and only if they act on disjoint qubit pairs, so the target interactions can be grouped into edge-coloring layers. 
The minimum number of such layers is the edge chromatic number $\chi'(G)$, which by Vizing's theorem satisfies $\chi'(G)\in\{d,d+1\}$.
In this sparse regime, the asymptotic circuit depth is set by the depth of the routing blocks, rather than by their  $N_\mathrm{perm}$. 
Using the notation of Sec.~\ref{subsec:general_setup}, $D_{\mathrm{route}}\sim\mathscr{D}(C)$, with $\mathscr{D}(C)=O(n)$ on a line and $\mathscr{D}(C)=O(\sqrt n)$ on a square grid.

We now specialize to RR graphs $G$ of degree $d=3$. 
They have $3n/2$ edges, edge density $3/(n-1)$, and $\chi'(G) \in \{3, 4\}$.
Since both $N_{\textrm{perm}} = \chi'(G)$ and $D_\mathrm{int}\leq \bar{d}$ are independent of $n$, the circuit depth in Eq.~\eqref{eq:circuit_depth} on line and square coupling maps scale as
\begin{equation}\label{eq:rr_depth}
    D_{\mathrm{line,\,RR}} \propto n,
    \qquad
    D_{\mathrm{grid,\,RR}} \propto \sqrt{n},
\end{equation}
respectively, as shown in Fig.~\ref{fig:circuit_metrics}(a,b).
Combining this with the relation $N \propto D \cdot O(n)$ for the number of entangling gate established in Sec.~\ref{subsec:general_setup} yields
\begin{equation}\label{eq:rr_gates}
    N_{\mathrm{line,\,RR}} \propto n^2,
    \qquad
    N_{\mathrm{grid,\,RR}} \propto n^{3/2},
\end{equation}
see Fig.~\ref{fig:circuit_metrics}(c).
The grid lowers each exponent by exactly $1/2$, and the gate count carries one further factor of $n$ relative to the depth.

For ER graphs with edge probability $q$, the number of target interactions scales as $O(qn^2)$, while the edge chromatic number $\chi'$, and hence the number of interaction layers, scales as $O(qn)$. 
Provided this factor does not saturate the routing advantage, the depth remains in the sparse-regime and inherits the $O(\sqrt n)$ improvement from the reduced grid diameter, leading to
\begin{equation}\label{eq:er_depth}
D_{\mathrm{line,\,ER}} \propto qn^2,
\qquad
D_{\mathrm{grid,\,ER}} \propto qn^{3/2},
\end{equation}
as shown in Fig.~\ref{fig:circuit_metrics}(d,e), and correspondingly
\begin{equation}\label{eq:er_gates}
N_{\mathrm{line,\,ER}} \propto qn^3,
\qquad
N_{\mathrm{grid,\,ER}} \propto qn^{5/2},
\end{equation}
see Fig.~\ref{fig:circuit_metrics}(f). 
Again, the grid lowers each exponent by $1/2$. 
We confirm the behavior of Eqs.~\eqref{eq:rr_depth}--\eqref{eq:er_gates} numerically in Sec.~\ref{sec:simulations}, where the \emph{line} and \emph{hybrid} strategies share the line-based exponents and the \emph{grid} and \emph{greedy} strategies share the grid-based ones.

\subsection{Dense regime}
\label{subsec:dense_regime}

We call the target graph dense when the number of qubit permutations $N_\mathrm{perm}$ in Eq.~\eqref{eq:circuit_depth} dominates the circuit depth. 
This happens when the $2|E|/(n\bar d)$ lower bound to $N_\mathrm{perm}$ of Sec.~\ref{sec:aspect_ratio} scales as $O(n)$ and is no longer small compared to the couplong map diameter $\mathscr{D}(C)$, see Fig.~\ref{fig:conceptual_graph_regimes}. 
Ref.~\cite{weidenfeller2202scaling} lower-bounds the two-qubit gate depth of the line-based and grid-based strategies by
\begin{equation}\label{eq:wf_bound_depth}
D_{\mathrm{line}, \mathrm{dense}} = \tfrac{3}{2}\,q\,n,
\qquad
D_{\mathrm{grid}, \mathrm{dense}} = \tfrac{7}{4}\,q\,n,
\end{equation}
where $q$ is the edge density.
Following the gate-count/depth scaling argument of Sec.~\ref{subsec:general_setup}, the corresponding entangling gate counts are
\begin{equation}\label{eq:wf_bound}
N_{\mathrm{line}, \mathrm{dense}} = \tfrac{3}{2}\,q\,n^{2},
\qquad
N_{\mathrm{grid}, \mathrm{dense}} = \tfrac{7}{4}\,q\,n^{2},
\end{equation}

\subsection{Bridging the two regimes}
\label{subsec:crossover}
\figureScalingGridWithDensity

The sparse and dense regimes correspond to different circuit complexity scalings resulting from different dominant contributions, see Fig.~\ref{fig:conceptual_graph_regimes}. 
In the sparse regime, $N_{\mathrm{perm}}$ is small and the routing distance $\mathscr{D}(C)$ dominates the circuit depth.
In the dense regime, $N_{\mathrm{perm}}$ scales as $O(n)$. 
The crossover between these limits occurs once the graph density is large enough that the permutation count $N_{\mathrm{perm}}$ saturates at its $O(n)$ ceiling. Past this point, added density no longer creates new permutation layers and instead packs more gates into each interaction layer, making the layers dense with $O(n \bar d)$ gates.

We can locate this boundary quantitatively through the average number of interactions per qubit, $qn$.
We do this by transpiling $e^{-i\gamma H_C}$ with $H_C$ based on MaxCut for ER graphs with size $n$ from $30$ to $120$ and density $q\in[0.08, 0.24]$ and fit the circuit depth to a model that we now describe.
The routing distance dominates while $qn$ remains small, and the gate count dominates once $qn$ grows large enough that most qubit pairs must interact.
Therefore, for a fixed density $q$, there exists a transition size $n^{\ast}$ below which the system is sparse and above which it is dense.
We thus model the circuit-depth by a function
\begin{equation}\label{eq:sat_limits}
D(n)\;\approx\;
\begin{cases}
b\,q \,n^{\alpha}, & n\ll n^{\ast},\\[4pt]
b\,q \,(n^{\ast})^{\,\alpha-\alpha_{\infty}}\,n^{\alpha_{\infty}}, & n\gg n^{\ast},
\end{cases}
\end{equation}
with two branches to represent both sparse and dense problems.
Here, $\alpha$ is the pre-saturation exponent predicted in Eq.~(\ref{eq:er_depth}) and $\alpha_{\infty}$ the saturated one from Eq.~(\ref{eq:wf_bound_depth}). 
ER graphs have $\alpha=2$ for the line and $\alpha=3/2$ for the grid strategies, while both saturate to $\alpha_{\infty}=1$.
To find $n^\ast$, we fit the single function
\begin{equation}\label{eq:sat}
D(n)=\frac{b\,q\,n^{\alpha}}{1+(n/n^{\ast})^{\,\alpha-\alpha_{\infty}}}
\end{equation}
to the depth of the transpiled cirucits.
Here, $b$ and $n^\ast$ are fit parameters, see dashed vertical lines in Fig.~\ref{fig:scaling_grid}(a-f).
Each routing strategy has a different $n^{*}$ reflecting its permutation efficiency.
A more efficient strategy realizes more edges of $H_C$ per permutation.
Therefore, at a fixed density, the more efficient strategies have a larger crossover size $n^{*}$. 
For example, \emph{hybrid} is more efficient than \emph{linear}, i.e., $n^{*}_\text{\emph{hybrid}}>n^{*}_\text{\emph{linear}}$ as shown in Fig.~\ref{fig:scaling_grid}, since it enables more interactions per qubit permutation with the same line SWAP strategy.
A divergent saturation size $n^{*}\to\infty$ at a given $q$ indicates that the considered graph sizes are sparse, e.g., \emph{grid} and \emph{greedy} at $q=8\%$, see Fig.~\ref{fig:scaling_grid}(a).
The grid-based strategies are more efficient than the line-based ones 
because their smaller diameter allows them to realize more interactions with the same amount of permutation layers, recall Fig.~\ref{fig:truncated}.
This explains why \emph{grid} and \emph{greedy} have larger $n^{*}$ values than \emph{linear} and \emph{hybrid}.

The local exponent
\begin{equation}\label{eq:alpha}
\hat\alpha(n)=\frac{\mathrm{d}\ln D}{\mathrm{d}\ln n},
\end{equation}
computed from the fits,
shows the two regimes since it measures the $\hat\alpha$ in $D\sim n^{\hat\alpha}$. 
A constant $\hat\alpha$ indicates that only one regime applies for the given $q$ value, see green line in Fig.~\ref{fig:scaling_grid}(g) and pink line in Fig.~\ref{fig:scaling_grid}(h).
Otherwise, $\alpha$ descends from its pre-saturation value $\alpha$ to the saturated value $\alpha_{\infty}$, tracking the transition between regimes.

We plot the boundary $n^{*}$ against $q$ for the \emph{linear}, \emph{hybrid}, and \emph{grid} strategies.
We observe a relation $qn^\ast(q)\sim\mathrm{const}$ that separates the sparse regime, below the curve, from the dense regime, above the curve, see Fig.~\ref{fig:saturation}.
We find boundaries $qn^\ast$ equal to $2.90 \pm 0.50$, $3.96\pm 0.59$, and $10.66\pm 0.30$ for \emph{linear}, \emph{hybrid}, and \emph{grid}, respectively.
A boundary further to the right is therefore preferable since the region below the boundary is larger.
In this region, shown in red in Fig.~\ref{fig:saturation}, the strategy produces shallower circuits than the \emph{linear} one.
Above the boundary, the corresponding strategy no longer benefits from the problem sparsity and can be substituted by the linear strategy which implements the full $O(n)$ permutation set optimally.
Therefore, ER graphs with $qn\leq 10.66$, i.e., below the boundary, should be transpiled with the \emph{grid} strategy.
If $qn>10.66$ then the \emph{linear} strategy should be used.

However, \emph{greedy} is expected to produce the rightmost boundary since for a given graph it produces circuits with at most as many layers as \emph{grid}. 
We do not include it in Fig.~\ref{fig:saturation}, as its $n^\ast$ values can exceed the maximum considered graph size of $n=120$.
When this happens the errors on the fit value of $n^\ast$ become too large, see Fig.~\ref{fig:scaling_grid}.
If the adaptive greedy search is too expensive to run, the \emph{grid} strategy is the natural fallback, since its boundary is the next furthest right.

\figureLinearScalingSaturationSizes

\setcounter{figure}{0}
\renewcommand{\thefigure}{D\arabic{figure}}
\figureCircuitMetricsAppendix
\figureMethodsComparisonAppendix

\section{Additional simulation results}\label{sec:other_parameters_results}

To confirm the circuit complexity scaling observed in Fig.~\ref{fig:circuit_metrics} we analyze degree $d=4$ RR graphs (edge density $\approx 4\%$) and ER graphs with edge probability $q=6.5\%$, see Fig.~\ref{fig:circuit_metrics_appendix}. 
Both are in the sparse regime.
As in Fig.~\ref{fig:circuit_metrics}, the \emph{linear} and \emph{hybrid} strategies have an identical asymptotic scaling due to their shared line-based SWAP structure. 
However, increasing $d$ from three to four increases the number of required qubit permutations which reduces the constant-factor advantage that the \emph{hybrid} strategy has over the \emph{linear} one.
As in Fig.~\ref{fig:circuit_metrics}, the \emph{grid} and \emph{greedy} strategies have a $O(\sqrt{n})$ advantage over line-based approaches in these sparse graphs, see App.~\ref{sec:asymptotic_analysis}.
The \emph{greedy} method continues to yield lower prefactors than the \emph{grid} strategy, consistent with its problem-adaptive routing.

We now compare the \emph{linear}, \emph{hybrid}, \emph{grid}, and \emph{greedy} methods to SABER based routing by transpiling the cost layer of QAOA for MC on ten random three-regular graphs with $n=100$ nodes. 
The SABER transpilation is done for a line, a heavy-hex~\cite{chamberland2020topological}, and a grid coupling map.
Across all ten instances the \emph{grid} and \emph{greedy} strategies require markedly fewer SWAP layers than the line-based strategies, with \emph{greedy} being the lowest throughout, see Fig.~\ref{fig:methods_comparison}(a).
The SABER methods are able to route the circuits with fewer SWAP and CZ gates than the SWAP strategy based networks.
Crucially, the SWAP strategy based methods substantially reduce the circuit depth relative to the SABRE-based methods, see Fig.~\ref{fig:methods_comparison}(c).
This implies that the SABER based routing methods leave many qubits idling in the resulting circuit.
Reducing the gate density of SWAP strategy based transpilation is left for future work.

\section{Practical Error Scaling Analysis}\label{sec:errors}

The experiments in this work are done on \emph{ibm\_miami}.
Its median CZ and single-qubit gate error rates are $p_\text{CZ} \approx 8.35 \times 10^{-3}$ and $p_{1q}\approx 2.21 \times 10^{-4}$, respectively.
Since $p_\text{CZ}>10p_\text{1q}$ we restrict the error analysis to two-qubit gates.
We now derive a simple threshold on the maximum number of two-qubit gates due to noise. 
We assume independent CZ errors with rate $p_\text{CZ} \sim 10^{-3}$ and independent measurement errors with rate $p_\text{meas} \sim 10^{-2}$ on $n$ qubits.
These numbers are slightly lower than those of \emph{ibm\_miami} to assess how our method scales as hardware error rates improve.
We estimate the success probability $p_{\mathrm{succ}}$ that no errors occur after $N_{\mathrm{CZ}}$ CZ gates and $n$ measurements by
\begin{equation}
    (1 - p_{\mathrm{CZ}})^{N_{\mathrm{CZ}}}(1 - p_{\mathrm{meas}})^n \approx e^{-N_{\mathrm{CZ}} p_{\mathrm{CZ}} - n p_{\mathrm{meas}}}.
\end{equation}
Imposing a minimum acceptable success probability $p_{\text{succ}, \min}$ bounds the number of CZ gates as
\begin{equation}
    N_{\mathrm{CZ}} \lesssim \frac{-\ln(p_{\text{succ}, \min}) - n p_{\mathrm{meas}}}{p_{\mathrm{CZ}}}.
\end{equation}

A $p_{\text{succ}, \min}$ of $10^{-2}$, retains a non-negligible number of correct samples out of the $10^4$ circuit repetitions. 
For $p_{\mathrm{CZ}} \sim 10^{-3}$ and $p_{\mathrm{meas}} \sim 10^{-2}$, this gives
\begin{equation}
    N_{\mathrm{CZ}} \sim 4.6\times 10^{3} - 10\,n,
\end{equation}
which for $n \in [80,100]$ yields $N_{\mathrm{CZ}} \in [3.6,3.8]\times 10^{3}$. 
We quote the smallest threshold, $\!3.6\times10^{3}$, as a conservative budget despite the optimistic error rates.

\section{Expected performance of random bitstring for MC}\label{sec:uniform_perf}

Finite-size and structural corrections increase the commonly quoted 50\% approximation ratio of uniform sampling for MC. 
For a uniform sampled bitstring {$x\in\{0,1\}^n$}, each vertex is independently assigned to one of the two partitions with probability $1/2$. 
Any edge $(u,v)\in E$ is cut if $x_u \neq x_v$, which occurs with probability $\mathbb{P}(x_u \neq x_v) = 1/2$. 
Hence, the expected cut value is $\mathbb{E}[f_{\mathrm{MC}}(x)] = |E|/2$ with an approximation ratio of
\begin{equation}
\mathbb{E}[r(x)] = \frac{|E|/2}{\mathrm{MC}(G)}.
\end{equation}
For random $d$-regular graphs the maximum cut satisfies $\mathrm{MC}(G)=|E|\big(1 - \Theta(1/\sqrt{d})\big)$ in the large-$n$ limit since only a vanishing fraction of edges cannot be simultaneously satisfied~\cite{dembo2017extremes, montanari2015optimization}. For $d=3$, this yields $\mathrm{MC}(G) \approx 0.92\,|E|$, and thus
\begin{equation}
\mathbb{E}[r(x)] \approx \frac{1}{2 \cdot 0.92} \approx 0.54.
\end{equation}

\section{Parameter optimization for MIS on ER graphs}\label{sec:param_optimization}
\setcounter{figure}{0}
\renewcommand{\thefigure}{G\arabic{figure}}
\figureParameterComparison

QAOA angle setting for MIS on ER graphs is less well characterized than for MC on RR graphs. 
Therefore, we rely on heuristic optimization methods. 
Since $p=1$ QAOA has two angles and its energy for quadratic cost Hamiltonians is efficient to simulate classically~\cite{Egger2021warmstartingquantum}, we can perform a grid search of the energy landscape to find near optimal angles.
We use a grid over $\beta, \gamma \in [0, \pi/2]$ with a $\pi/100$ step size.
We refine the resulting near optimal angles with $40$ iterations of the COBYLA optimizer from SciPy with an initial trust-region radius of $0.05$.
The implementation leverages the \texttt{DepthOneScanTrainer}, and the \texttt{ScipyTrainer} coupled with the \texttt{EfficientDepthOneEvaluator} from the QAOA Training Pipeline~\cite{qaoa_training_pipeline}. 
This two-stage procedure balances global exploration of the energy landscape with efficient local convergence.

We optimize the $p=2$ angles with COBYLA set to an initial trust region of 0.2.
The energy is evaluated by a MPS-based simulator with bond dimension $64$ and a truncation threshold of $10^{-10}$.
The initial point given to COBYLA is based on the MC fixed angles of Ref.~\cite{wurtz2021fixed}.
For a given ER graph $G$, we use the fixed-angles of the RR graph with degree closest to the average node degree of $G$.
The $\gamma_1$ and $\gamma_2$ of these fixed angles are then rescaled following Eq.~(8) of Ref.~\cite{sureshbabu2024parameter} to account for the fact the Pauli terms in the MIS Hamiltonian differ from those in MC.
The resulting $\beta$ and rescaled $\gamma$ angles are the initial point given to COBYLA.

Concretely, the scaling factor combines the root-mean-square coefficients
of the single-$Z$ and $ZZ$ terms,
\[
\lambda = \left( \frac{1}{n_Z}\sum_{i} w_i^2
  + \frac{1}{n_{ZZ}}\sum_{(i,j)} w_{ij}^2 \right)^{1/2},
\]
with $n_Z$ and $n_{ZZ}$ the numbers of single-$Z$ and $ZZ$ terms and coefficients $w_i$ and $w_{ij}$. 
Our Hamiltonian $H_{\textrm{MIS}}$ resulting from Eq.~\eqref{eq:cost_function_MIS} 
has $w_i=\frac{1}{2}(\deg_i-1)$ and $w_{ij}=\frac{1}{2}$, leading to
\begin{align}
\lambda
&= \left( \frac{1}{n_Z}\sum_{i}
  \Big(\frac{\deg_i-1}{2}\Big)^2 + \frac{1}{4} \right)^{1/2}
  \label{eq:lambda_mis_def} \\
&= \frac{1}{2}\sqrt{\,1 + \big(\mathbb{E}(\deg_i)-1\big)^2
  + \mathrm{Var}(\deg_i)\,}.
  \label{eq:lambda_mis_moments}
\end{align}
The angles then transfer as
\begin{equation}
\boldsymbol{\gamma} = \frac{1}{2\lambda}\boldsymbol{\gamma}_{\mathrm{base}},
\qquad
\boldsymbol{\beta} = \boldsymbol{\beta}_{\mathrm{base}},
\end{equation}
where the factor $1/2$ absorbs the normalization convention of our cost Hamiltonian relative to the reference. This transfer avoids costly instance-specific training but stays approximate for individual ER instances, in particular under broad degree fluctuations.

Ref.~\cite{sureshbabu2024parameter} calibrates on RR graphs, where $\mathrm{Var}(\deg_i)=0$ and thus $\lambda$ depends only on the mean degree.
On ER graphs, the nonzero variance signals a heterogeneous structure absent from the reference, so the transfer crosses graph families and is not guaranteed to be near-optimal.
However, we find that this procedure produces better angles than initializing COBYLA either by linearly interpolating the optimal $p=1$ angles $(\beta_1^\star,\gamma_1^\star)$~\cite{zhou2020quantum} or by trivially extending them to the initial point $(\beta_1^\star, 0, \gamma_1^\star, 0)$, see Fig.~\ref{fig:parameter_comparison}.

\section{Experimental Post-processing effects}\label{sec:experimental_postprocessing}
\setcounter{figure}{0}
\renewcommand{\thefigure}{H\arabic{figure}}

The hardware results in the main text are based on all  quantum samples. 
Post-selection can improve the quality of the final outcome at the cost of retaining only a fraction of the measured samples.
Here, we describe and test a hardware-based post-selection scheme implemented using the \texttt{noise\_management.post\_selection} module from the Qiskit Addon Utils~\cite{qiskit-addon-utils}.

The post-selection scheme we use targets readout and relaxation errors that occur at the end of the circuit. 
For each monitored qubit $i$, we perform an inversion test. 
After a first measurement we apply an $X$ gate, to flip the qubit, and then measure again, see Fig.~\ref{fig:postselection_circuit}. 
In the absence of measurement-stage errors, the two outcomes $m_i$ and $m_i'$ must be anticorrelated. 
A shot is thus discarded if the two outcomes agree for any one qubit, i.e. there is a qubit $i$ for which $m_i = m_i'$. 
This test is independent of the ideal output distribution and can be applied uniformly across qubits and problem instances.
The retained shots are then used to compute the mean approximation ratio.

\figurePostselectionCircuitAppendix
\figurePostselectionAppendix

Post-selection produces modest improvements to the mean approximation ratio, which is why we report raw samples in the main text, compare darker curves relative to the lighter ones in Fig.~\ref{fig:postselection}. 
For the MC problem on RR3 graphs at $p=1$, the improvement is consistent for both the \emph{linear} and \emph{greedy} strategies, but reaches at most $2.76\%$ and $2.27\%$, respectively, at 48 qubits, see Fig.~\ref{fig:postselection}(a).
For $p=2$, it persists up to $n=64$ qubits, see Fig.~\ref{fig:postselection}(b). 
At larger sizes, the circuits become sufficiently deep that too few samples survive post-selection for reliable statistics. For example, out of $10,000$ shots at $n=72$ and $p=2$, only 6 survive on average for the \emph{linear} strategy and 16 for the \emph{greedy} strategy, see Fig.~\ref{fig:postselection}(c).

A similar trend is observed for the MIS problem on ER graphs. At $p=1$, post-selection consistently improves the results (except for the \emph{greedy} strategy at $n=72$), see Fig.~\ref{fig:postselection}(d). At $p=2$, the improvement remains visible only while the hardware performance is statistically distinguishable from uniform random sampling, namely up to $n=48$, see Fig.~\ref{fig:postselection}(e). The effect of post-selection is again more pronounced for circuits transpiled with the \emph{linear} strategy, for which only a fraction $0.95\%$ of shots is retained at $p=2$, compared with $1.12\%$  for the \emph{greedy} strategy, see Fig.~\ref{fig:postselection}(f).

\newpage
\bibliography{refs}

\end{document}

%% file: allcommands.tex
\usepackage{algorithm}
\usepackage{algpseudocode}


\usepackage[dvipsnames]{xcolor,colortbl}
\usepackage{xspace,tabularx}
\usepackage{mathtools,amsthm,amssymb, amsmath} 

\usepackage{enumitem} 
\setenumerate[0]{label={\normalfont (\roman*)}}
\usepackage{graphicx}
\usepackage{subcaption}
\usepackage{ragged2e}
\usepackage{booktabs}
\usepackage{mathrsfs}

\usepackage{tikz}
\usetikzlibrary{quantikz2}

\usepackage[T1]{fontenc}
\usepackage[UKenglish]{babel}
\usepackage[scaled=0.86]{helvet}
\DeclareMathAlphabet{\mathcal}{OMS}{ntxm}{m}{n}
\usepackage{dsfont} 
\let\mathbb\relax
\let\mathbb\mathds
\usepackage{stmaryrd} 

\usepackage{dcolumn}
\usepackage{bm}

\definecolor{ibmblue}{HTML}{0043ce}
\usepackage[colorlinks=true,urlcolor=ibmblue,citecolor=ibmblue,linkcolor=ibmblue]{hyperref}
\usepackage{url}
\usepackage[nameinlink,capitalize,noabbrev]{cleveref}
\crefname{enumi}{Step}{Steps}

\usepackage{physics}

%% file: figures.tex
\newcommand{\figureAllLayers}{
    \begin{figure}[!htb]
    \centering
    \includegraphics[width=\linewidth]{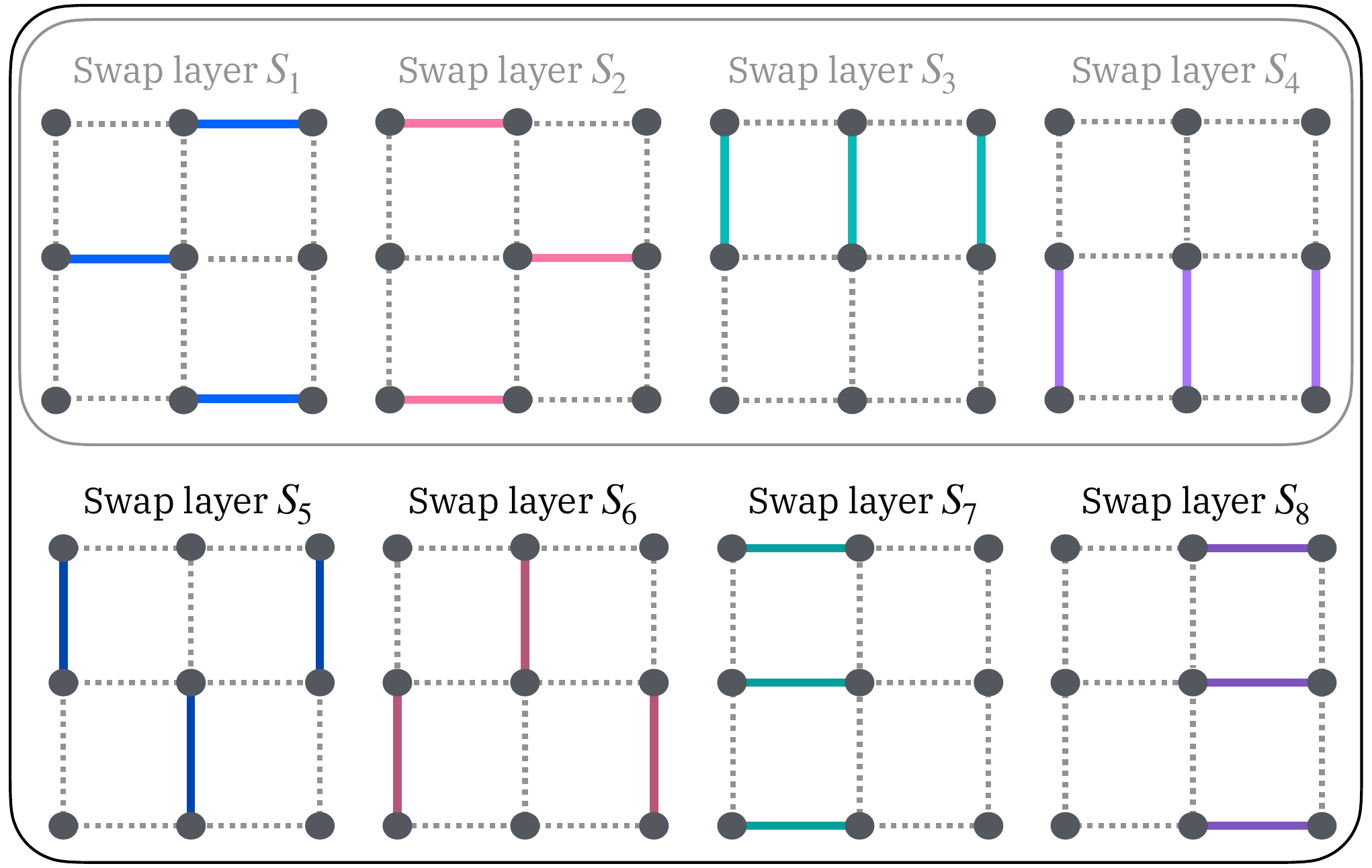} 
    \caption[singlelinecheck=false]{\justifying Example of two basis sets of SWAP layers for a $3\times 3$ grid: $\mathcal{B}_{\mathrm{grid}}$ (top row) and its extended version $\mathcal{B}_{\mathrm{grid}}^{\mathrm{extended}}$ (all).
    Here, each dot is a qubit, and the dashed lines show the coupling map.
    The highlighted edges in each SWAP layer show SWAP gates between qubits.
    } 
    \label{fig:all_layers}
\end{figure}
}

\newcommand{\figureLinearAndHybrid}{
    \begin{figure}[!htbp]
        \centering
        \includegraphics[width=\columnwidth]{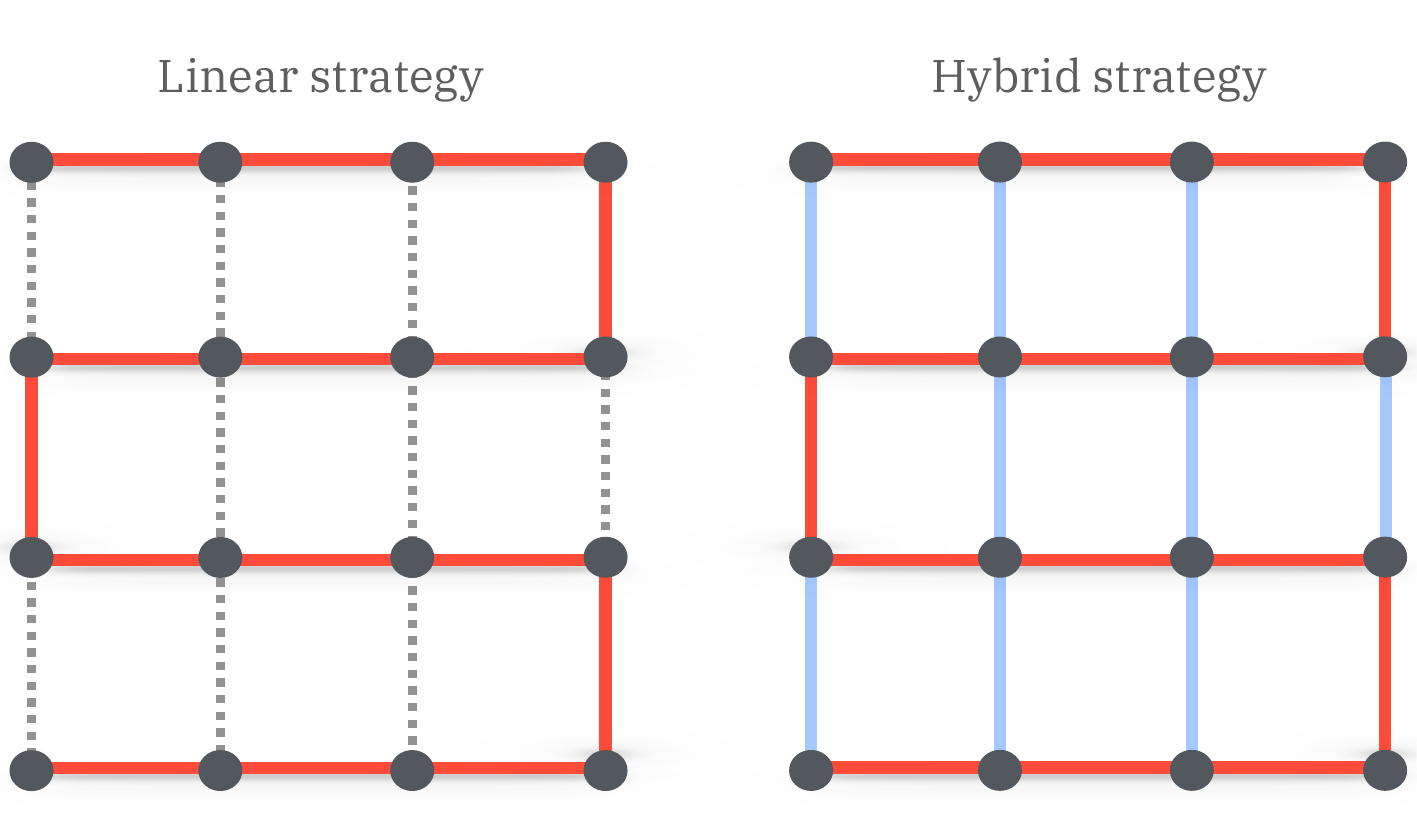}
        \caption{\justifying \emph{Linear} and \emph{hybrid} strategy on a $4\times 4$ rectangular grid coupling map.
        Both strategies route along the longest line of qubits (red), but the \emph{hybrid} strategy can additionally apply interactions on the remaining edges of the coupling map (light blue).
    }
        \label{fig:fig_hybrid}
    \end{figure}
}

\newcommand{\figureAlgo}{
    \begin{figure}[htbp]
    \centering
    \includegraphics[width=\linewidth]{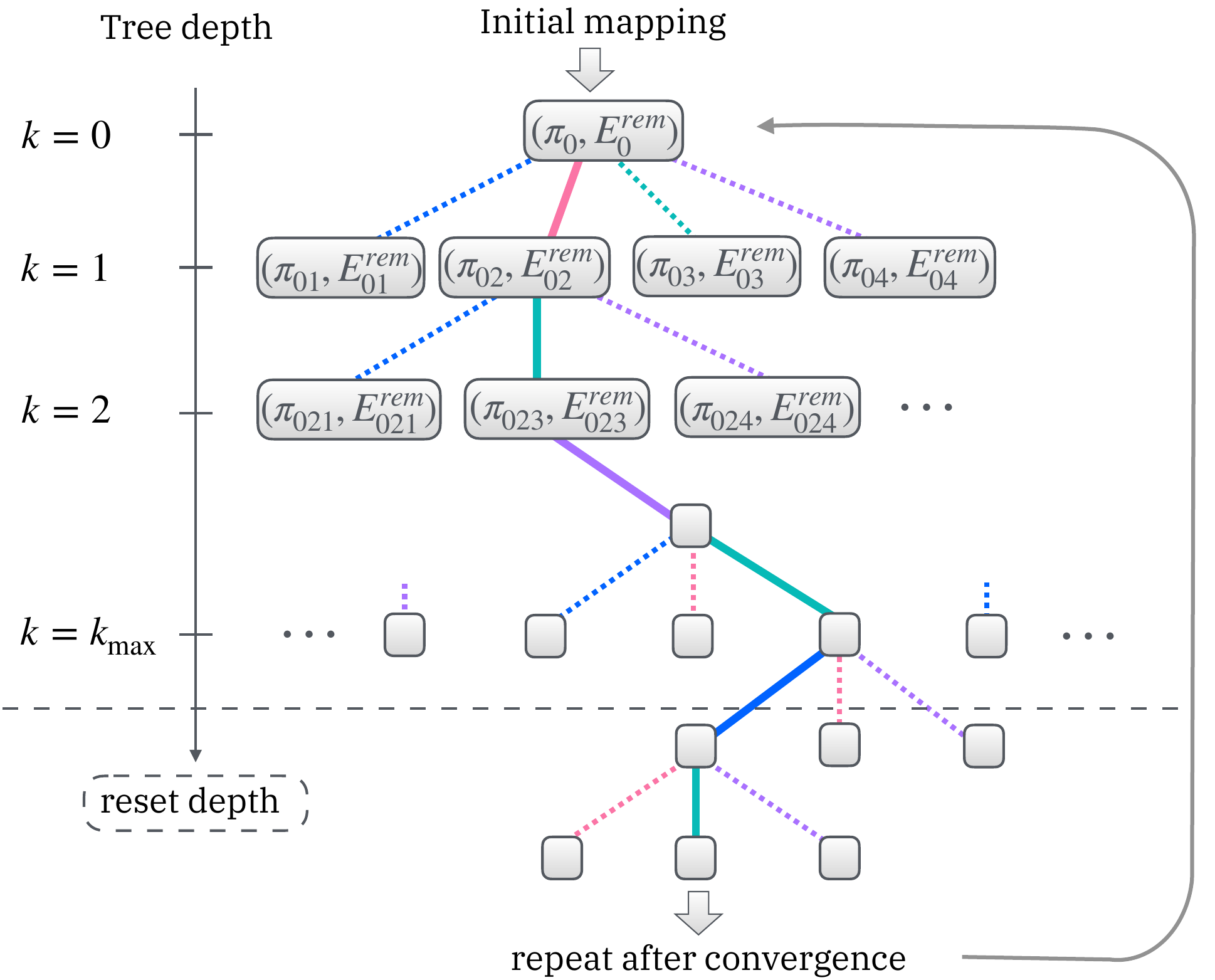} 
    \caption[singlelinecheck=false]{\justifying 
    \emph{Greedy} searches a tree of SWAP layer sequences.
    The edges correspond to SWAP layers and are colored according to the basis $\mathcal{B}_{\text{grid}}$ shown in Fig.~\ref{fig:all_layers}. 
    The nodes encode qubit layouts $\pi$ and the remaining interactions $E^{\textrm{rem}}$. 
    At depth $k=1$, possible layouts are denoted by $\pi_{01}, \pi_{02}, \pi_{03}, \pi_{04}$, corresponding to their respective SWAP sequences. 
    The tree is explored up to a fixed depth $k_{\max}$, after which a leaf minimizing the number of remaining edges is selected to continue the search until all interactions are implemented, yielding $\mathcal{S}_{\text{greedy}}$.
    } 
    \label{fig:algo}
\end{figure}
}

\newcommand{\figureCircuitMetrics}{
\begin{figure*}[htbp]
    \centering

    \includegraphics[width=\linewidth]{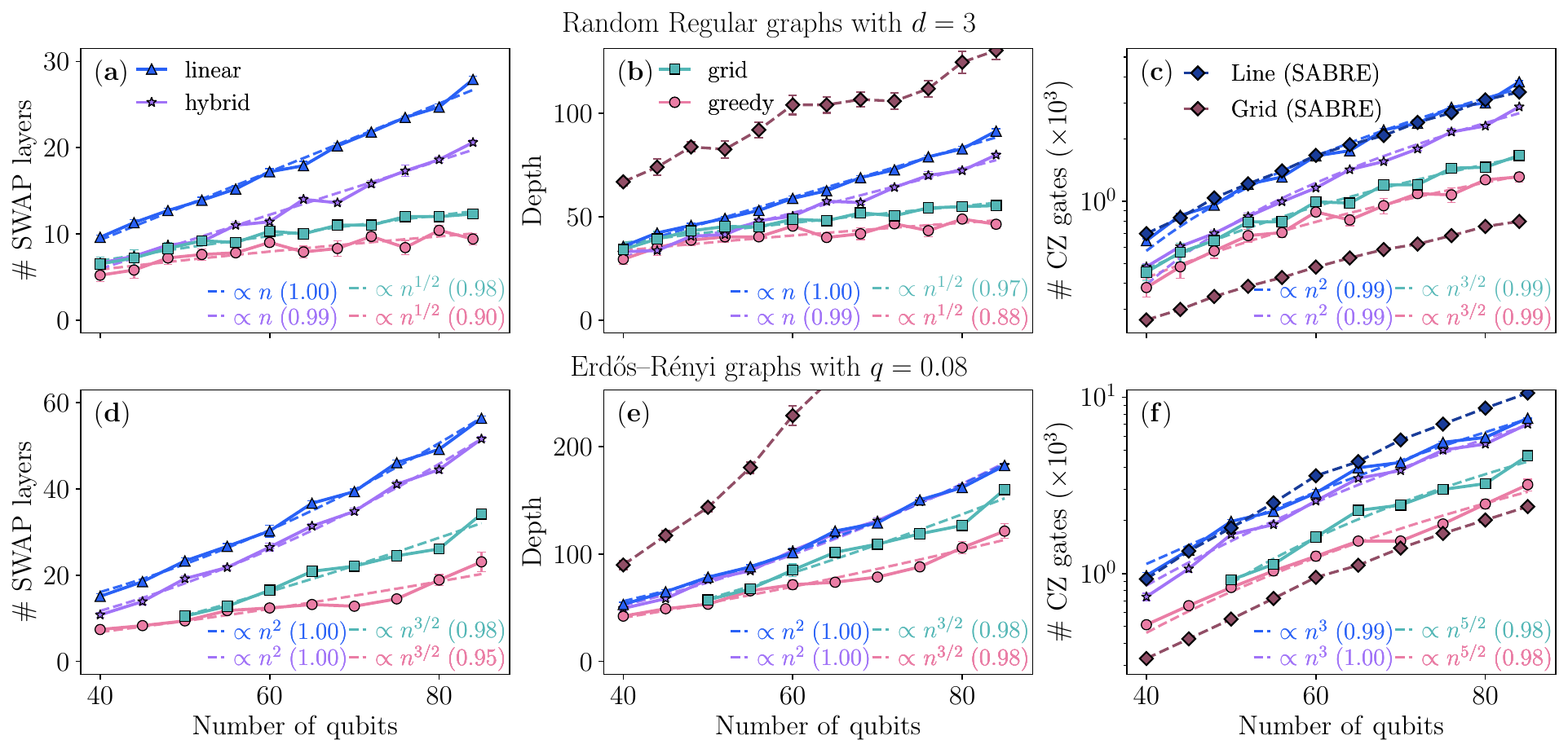}
    
    \caption{\justifying Circuit complexity versus graph size of $e^{-i\gamma H_C}$ transpiled to a rectangular grid with different strategies. The \emph{linear} strategy is the current state-of-the-art and is based on $(\mathcal{S}_{\mathrm{line}}, C_\mathrm{grid}^\mathrm{line})$; the \emph{hybrid} one is our first contribution and is based on $(\mathcal{S}_{\mathrm{line}}, C_{\mathrm{grid}})$; the \emph{grid} strategy uses $(\mathcal{S}_{\mathrm{grid}}, C_{\mathrm{grid}})$~\cite{weidenfeller2202scaling}; the \emph{greedy} strategy is our main contribution and uses $(\mathcal{S}_{\mathrm{grid}}^{\mathrm{greedy}}, C_{\mathrm{grid}})$; 
    the Line (SABRE) and Grid (SABRE) methods use the out-of-the-box Qiskit transpiler~\cite{li2019mapping, zou2024lightsabre, javadi2024quantum} on $C_\mathrm{grid}^\mathrm{line}$ and $C_\mathrm{grid}$, respectively. 
    The Line (SABRE) results are omitted from panels (b) and (e) because they fall outside the plotted range.
    Scalings are derived in App.~\ref{sec:asymptotic_analysis}. Fits are performed on averaged $y$ values over 10 instances; values in parentheses denote the coefficients of determination $R$.
    The $y$-axis upper bounds in (a) and (b) are half those in (d) and (e), respectively, reflecting the approximately halved number of edges in these graphs. 
    }
    \label{fig:circuit_metrics}
\end{figure*}
}

\newcommand{\figureHeatmaps}{
\begin{figure*}[htbp]
    \centering
    \includegraphics[width=\linewidth]{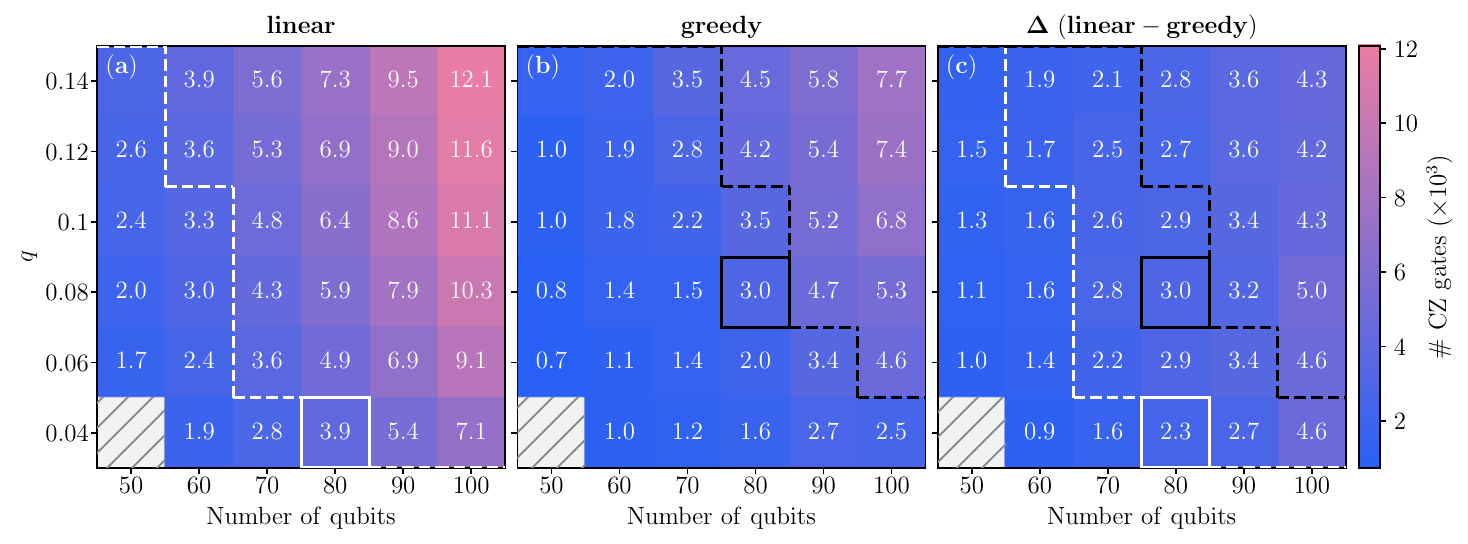}
    \caption{\justifying Entangling gates scaling of $e^{-i\gamma H_C}$ transpiled ton a grid with the \emph{linear} (a) and \emph{greedy} (b) SWAP strategies versus the graph size and edge density of the underlying Erd\H{o}s--R\'enyi graphs.
    Each data point is the average of ten instances.
    The colors and numbers show the number of CZ gates.
    White and black dashed lines indicate the estimated $\sim 3{,}600$ CZ gates feasibility thresholds for \emph{linear} and \emph{greedy} strategies, respectively, which assumes optimistic error rates, not those of the hardware device.
    The region between them in (c) marks instances enabled by our approach. 
    Graphs highlighted by the white square have a similar density as the RR graphs in the hardware experiments from Fig.~\ref{fig:hardware_results_all}(a-d). 
    The black square corresponds to the exact same circuits as the hardware experiments from Fig.~\ref{fig:hardware_results_all}(e-h). 
    }
    \label{fig:heatmaps}
\end{figure*}
}

\newcommand{\figureTruncatedImplementations}{
\begin{figure}[htbp]
    \centering
    \includegraphics[width=\linewidth]{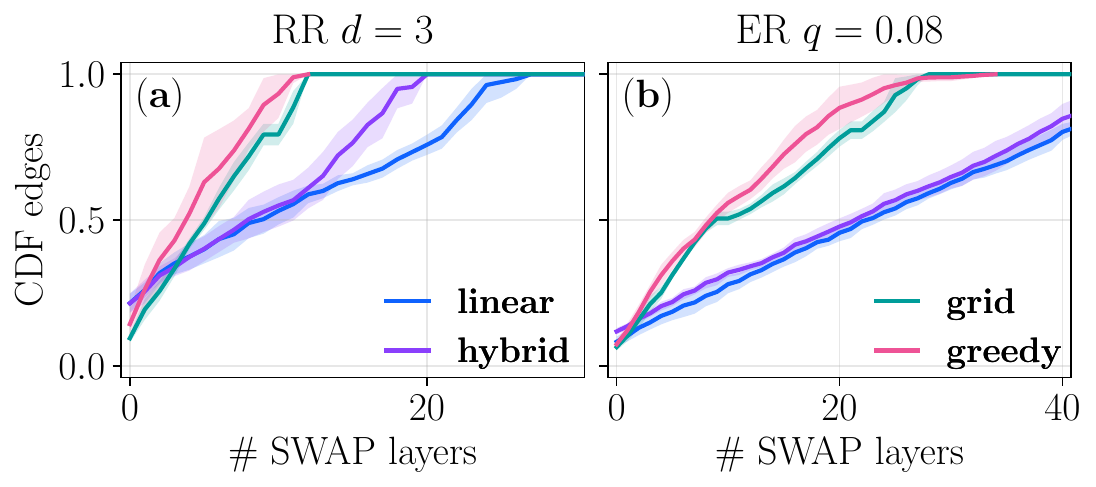}
    \caption{ \justifying
    Truncation quality, measured by the fraction of implementable target interactions, is shown as a function of the number of SWAP layers for different transpilation strategies. Curves shifted to the left indicate higher performance. All graphs contain $80$ nodes, and the shaded regions denote the variation across 10 graph instances.
    }
    \label{fig:truncated}
\end{figure}
}

\newcommand{\figureHardwareResultsAll}{
\begin{figure*}[htbp]
    
    \centering
    \includegraphics[width=\linewidth]{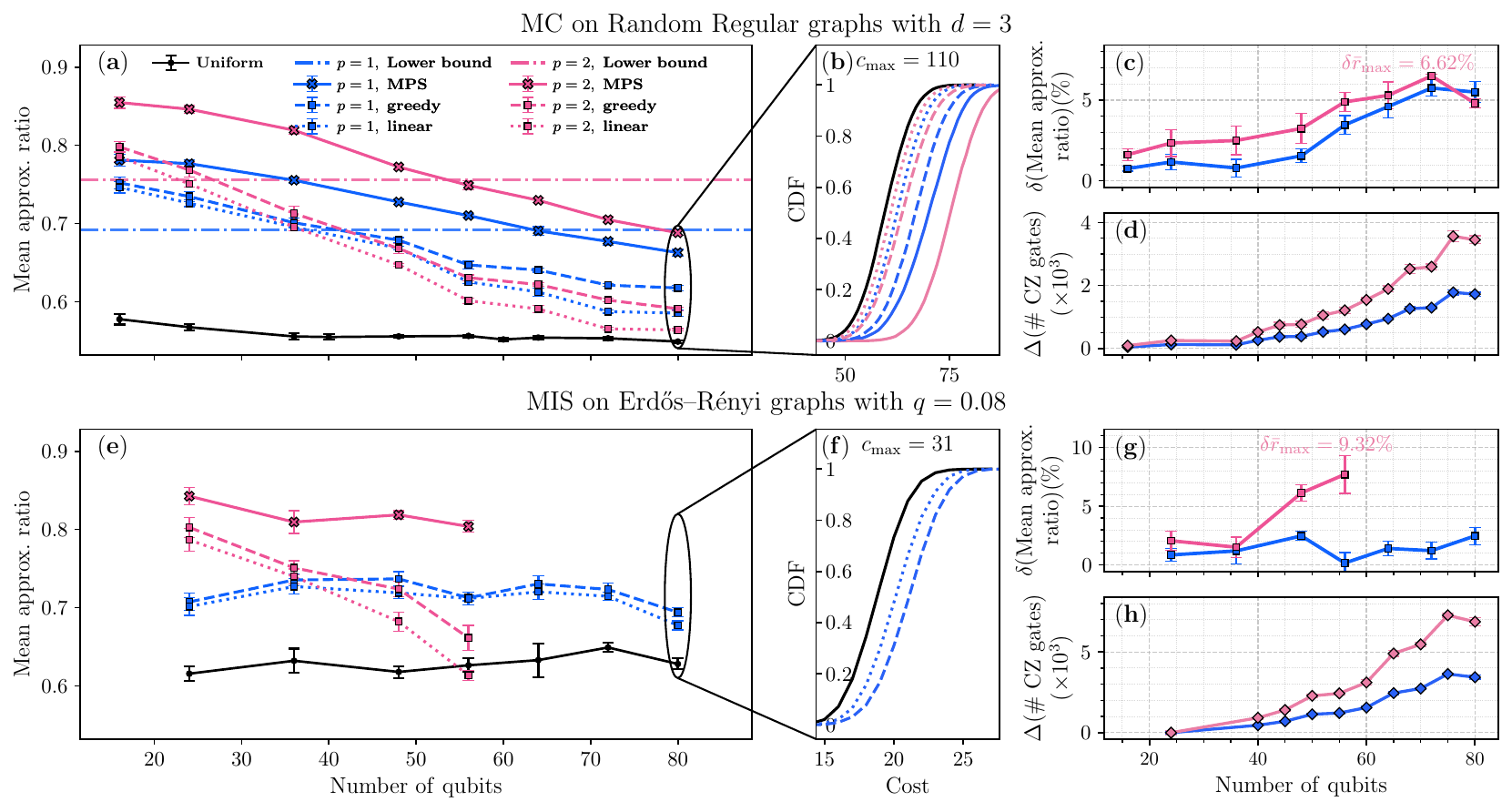}
    \caption{\justifying Hardware performance analysis of QAOA circuits with $p=1$ (blue) and $p=2$ (pink) layers, transpiled using \emph{linear} (dotted) and \emph{greedy} (dashed) strategies. 
    The theoretical guarantee bounds in (a) are taken from~\cite{wurtz2021fixed}. The MPS solid curves can underestimate the true noiseless QAOA value due to finite bond dimension (64) truncation, especially at larger system sizes. 
    Results are averaged over $10$ graph instances, with $10{,}000$ samples per circuit. No error mitigation or post-selection is applied.
    }
    \label{fig:hardware_results_all}
\end{figure*}
}

\newcommand{\figureHyperparameters}{
\begin{figure}[htbp]
    \centering

    \begin{subfigure}[t]{\linewidth}
        \centering
        \includegraphics[width=\linewidth]{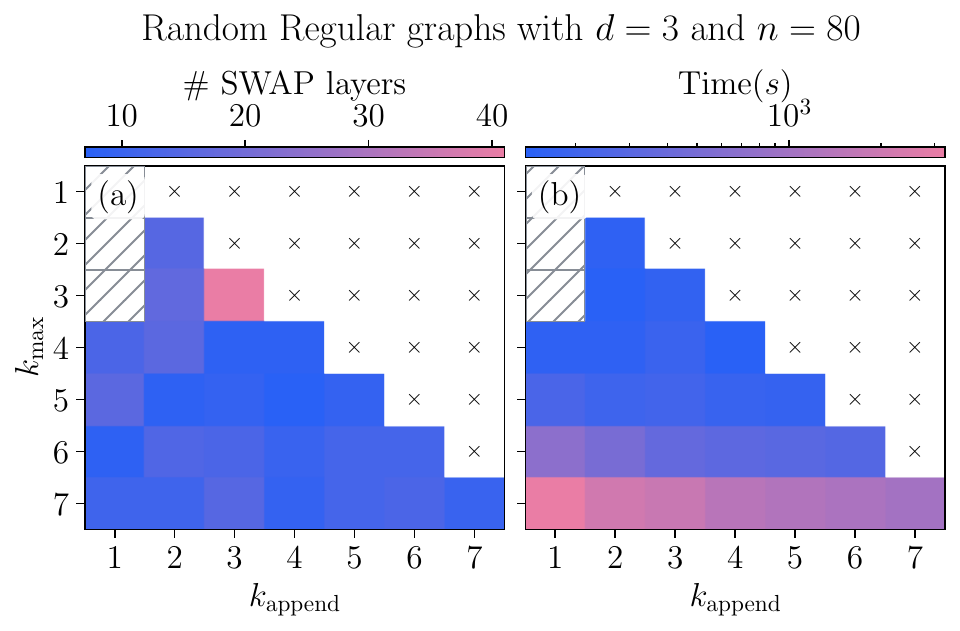}
    \end{subfigure}
    \hfill
    \begin{subfigure}[t]{\linewidth}
        \centering
        \includegraphics[width=\linewidth]{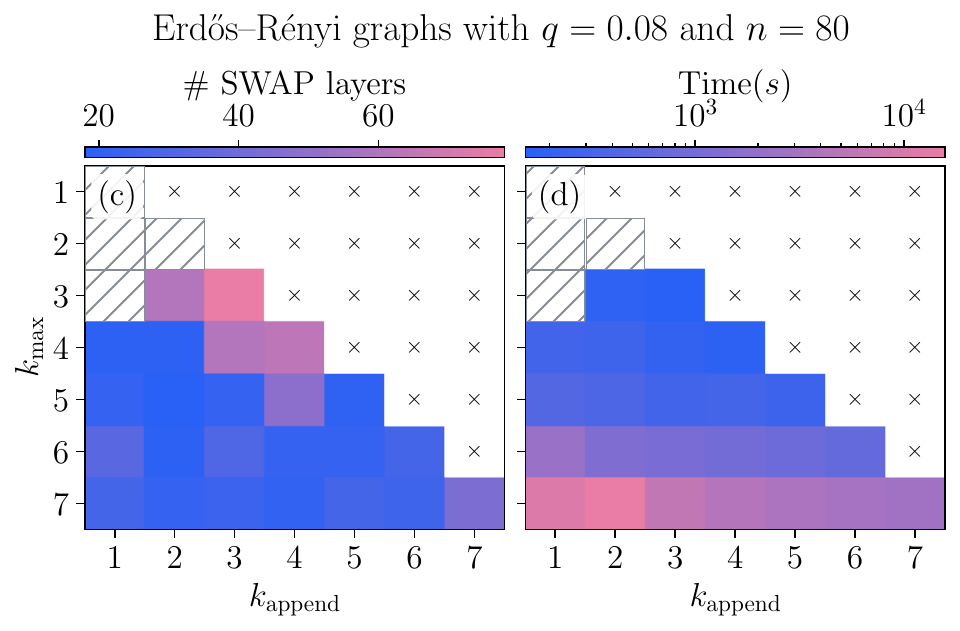}
    \end{subfigure}
    
    \caption{ \justifying
    Parameter sweep analyzing the performance-time trade-off of our \emph{greedy} strategy. Hatched squares indicate instances where no solution was found, while crossed squares denote invalid parameter configurations. Data points are averaged over 10 instances.
    }
    \label{fig:hyperparams}
\end{figure}
}

\newcommand{\figureSquarenessImpactAppendix}{
    \begin{figure*}[htbp]
    \centering
    \includegraphics[width=\linewidth]{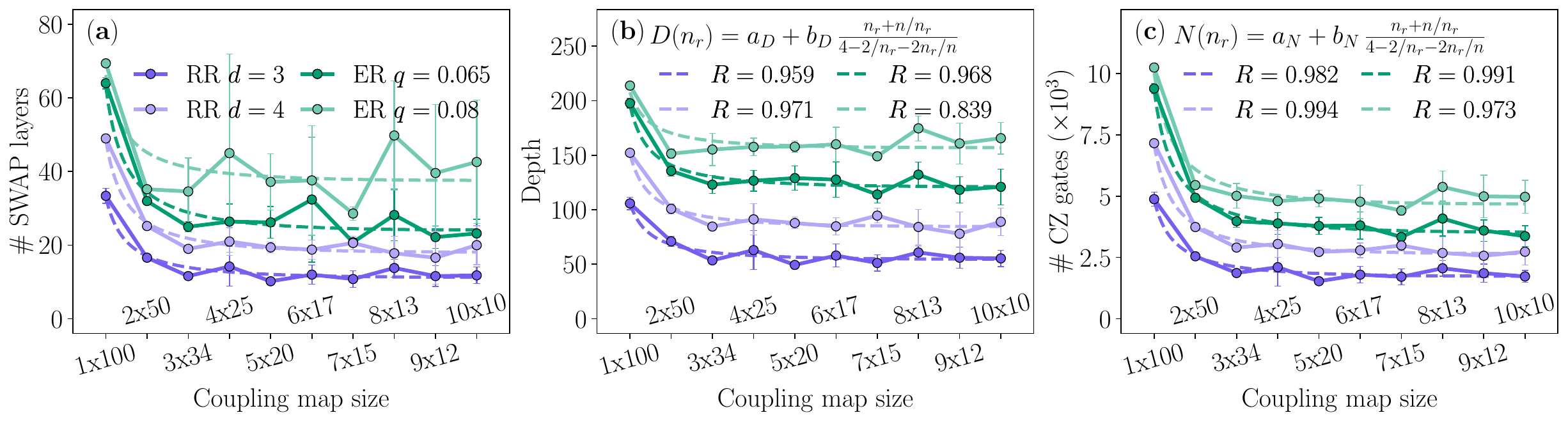} 
    \caption[singlelinecheck=false]{ \justifying
    Circuit complexity versus the rectangular lattice aspect ratio for $e^{-i\gamma H_C}$ defined on graphs with $100$ nodes and transpiled with the \emph{greedy} strategy. 
    The coupling map has size $n_r\times n_c$, with $n_r$ rows and $n_c$ columns. 
    Each curve corresponds to a different graph family, as indicated in the legend of panel (a): random-regular graphs with degrees three and four (purple), and Erdős--Rényi graphs with edge probabilities $0.065$ and $0.08$ (green). 
    Dashed curves show fits to Eq.~\eqref{eq:layer_depth} in panels (a,b) and to Eq.~\eqref{eq:layer_volume} in panel (c). 
    Data points are averaged over $10$ graph instances. 
    The coefficient of determination $R$ reported in panels (b,c) corresponds to the fit to the averaged data.
    }
    \label{fig:squareness_impact}
\end{figure*}
}

\newcommand{\figureConceptualGraphRegimes}{
    \begin{figure*}[t]
      \centering
      \includegraphics[width=\textwidth]{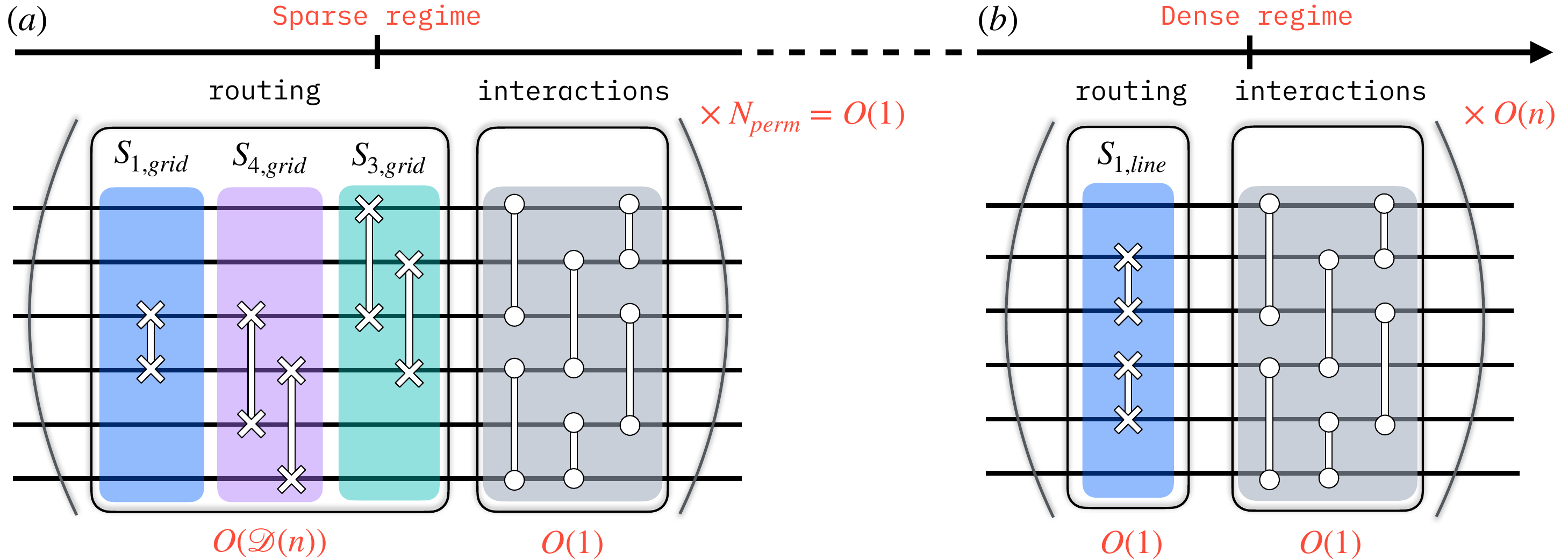}
      \caption{\justifying
    Schematic illustration of the sparse and dense graph regimes, defined by the dominant contribution to circuit-depth scaling. The key quantity is the number of qubit permutations, $N_{\mathrm{perm}}$, required to implement all target interactions. 
    In the sparse regime, only $N_{\mathrm{perm}}=O(1)$ routing blocks are required, so the depth is dominated by the routing, whose depth scales with the diameter $\mathscr{D}(n)$ of an $n-$node graph inheriting the hardware's connectivity, e.g. $O(\sqrt n)$ for a two-dimensional grid.
    In the dense regime, the benefit of additional qubit permutations is saturated, i.e. $N_{\mathrm{perm}}=O(n)$, so the depth scales linearly, with its prefactor governed by the depth of the interaction layers. The circuits shown are representative examples; the routing sequence may differ between blocks.
      }
      \label{fig:conceptual_graph_regimes}
    \end{figure*}
}

\newcommand{\figureScalingGridWithDensity}{
    \begin{figure*}[t]
      \centering
      \includegraphics[width=\textwidth]{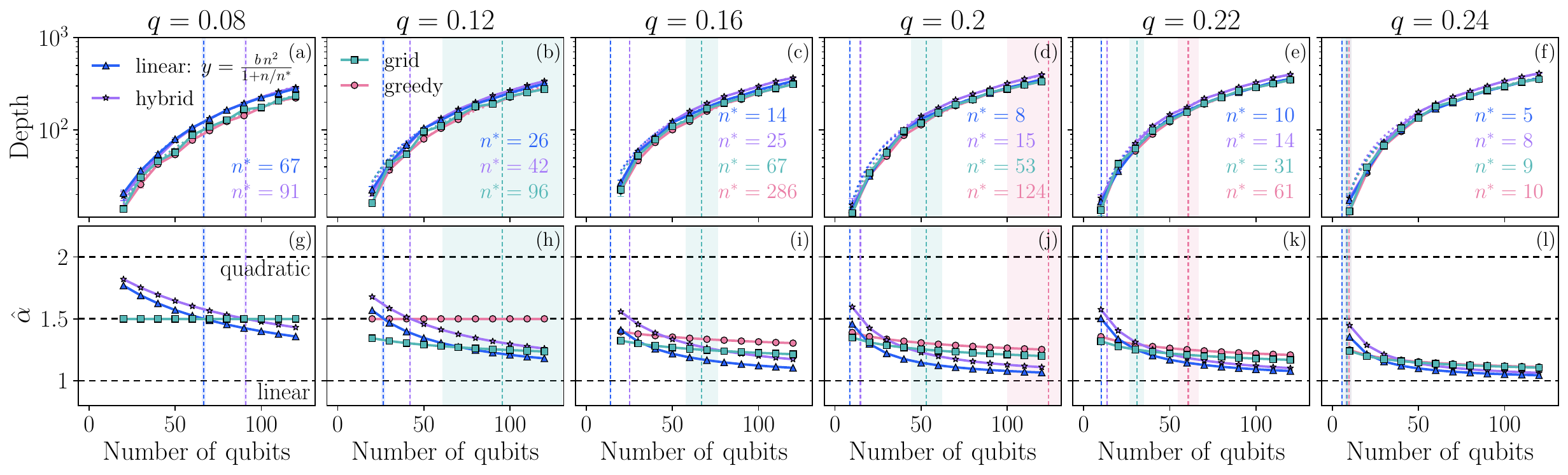}
      \caption{\justifying
      Scaling of depth and its local exponent $\hat \alpha(n) = d \ln D / d \ln n$ versus problem size $n$ for ER graphs at various densities $q$ (columns), for the \emph{line}, \emph{hybrid}, \emph{grid}, and \emph{greedy} strategies. 
      Dashed vertical lines show the fitted $n^{\ast}$ from Eq.~\eqref{eq:sat}, also written in the bottom legends. 
      When omitted, it means $n^{\ast}\to\infty$, corresponding to an encompassing sparse regime. 
      All coefficients of determination satisfy $R\geq 0.98$.
      Datapoints are averaged over ten graph instances.
      }
      \label{fig:scaling_grid}
    \end{figure*}
}

\newcommand{\figureLinearScalingSaturationSizes}{
    \begin{figure}[t]
      \centering
      \includegraphics[width=\columnwidth]{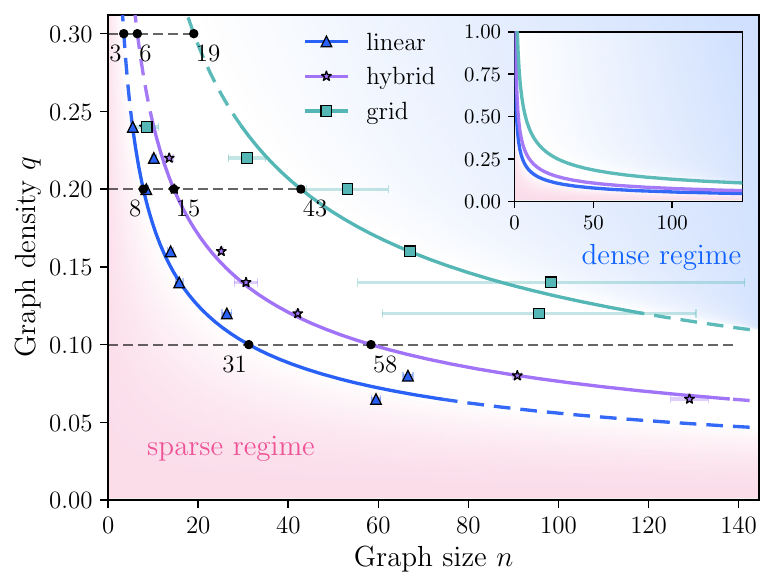}
      \caption{\justifying
      Boundaries for each routing strategy, separating the sparse from the dense regime as a function of graph size $n$ (horizontal axis) and density $q$ (vertical axis). 
      Markers denote the fitted saturation sizes $n^{\ast}$ from Eq.~\eqref{eq:sat}. 
      Solid (dashed) curves lie inside (outside) the measured density range.
      }
      \label{fig:saturation}
    \end{figure}
}

\newcommand{\figureCircuitMetricsAppendix}{
\begin{figure*}[htbp]
    \centering

    \includegraphics[width=\linewidth]{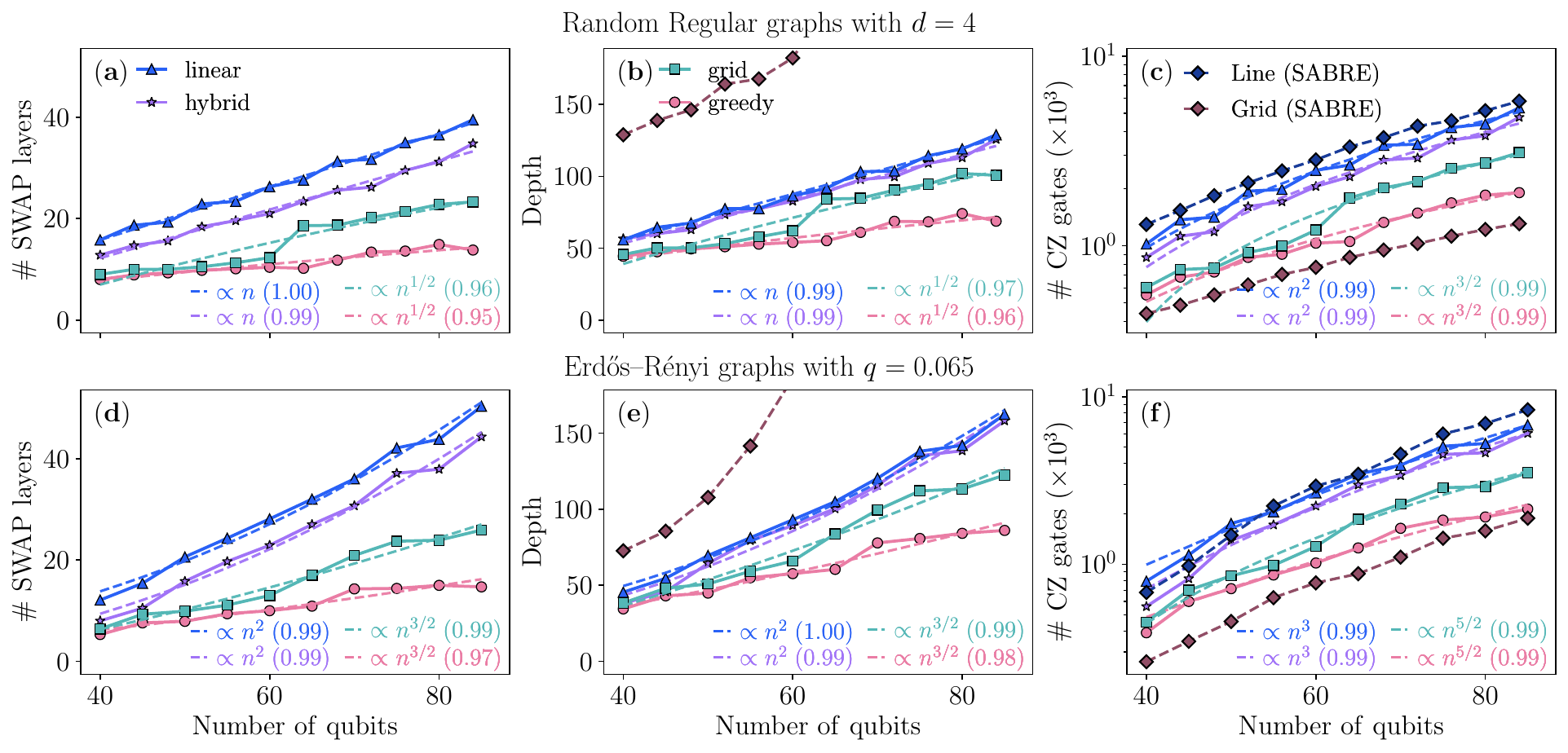}

    \caption[singlelinecheck=false]{
    \justifying Additional results showing the circuit complexity versus graph size for time-evolution circuits on a rectangular lattice under different transpilation strategies. 
     Scalings are derived in App.~\ref{sec:asymptotic_analysis}. Fits are performed on averaged $y$ values over 10 instances; values in parentheses denote the coefficients of determination $R$ of the fits.
    The $y$-axis upper bounds in (a), (b), and (c) are the same as those in (d), (e) and (f), respectively, reflecting the approximately same number of edges in these graphs. 
    }
    \label{fig:circuit_metrics_appendix}
\end{figure*}
}

\newcommand{\figureMethodsComparisonAppendix}{
    \begin{figure*}[htbp]
    \centering
    \includegraphics[width=\linewidth]{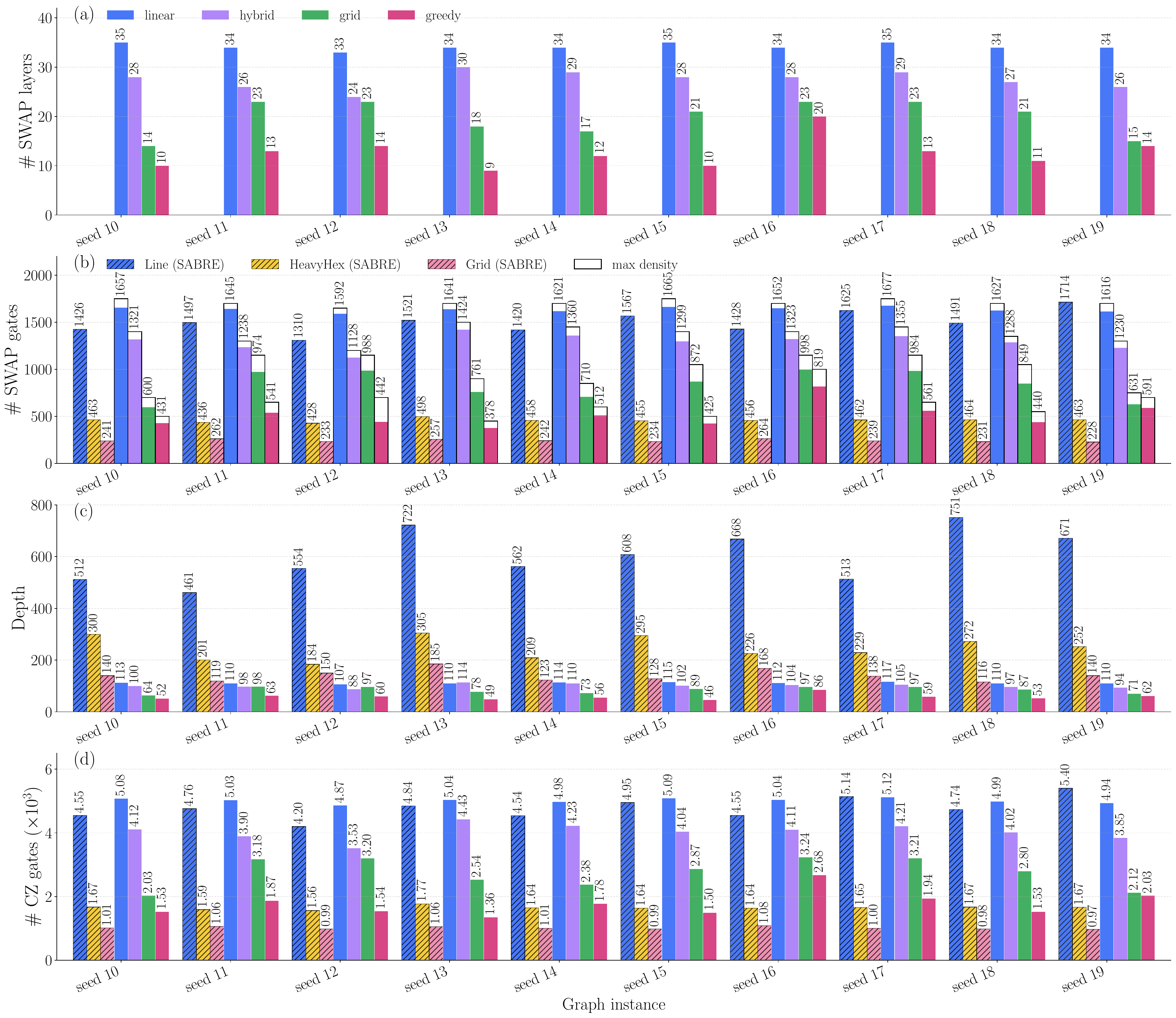} 
    \caption[singlelinecheck=false]{
    \justifying 
    Per-instance circuit metrics for depth-one QAOA cost layers
    $e^{-i\gamma H_C}$ of random three-regular graphs with $n=100$ nodes. Each group on the horizontal axis is one graph instance, labeled by its generating seed, so the bars are single-instance values without averaging.
    Panel (a) compares the \emph{linear}, \emph{hybrid}, \emph{grid}, and \emph{greedy} SWAP strategies in terms of number of SWAP layers, while the other panels also show the performance of the out-of-the-box Qiskit transpiler with SABRE routing~\cite{li2019mapping, zou2024lightsabre, javadi2024quantum} on the line, heavy-hex, and grid coupling maps.
    Relevant comparisons look at strategies that either i) belong to the same family, i.e., SABRE or SWAP strategies, or ii) use the same coupling map, i.e. \emph{linear} versus Line (SABRE), and any other SWAP strategy (\emph{hybrid}, \emph{grid}, or \emph{greedy}) versus Grid (SABRE).
    The maximum-density reference in panel (b) shows the value of the number of SWAP layers multiplied by a factor of $50$, corresponding to a maximally-dense SWAP layer.
    }
    \label{fig:methods_comparison}
\end{figure*}
}

\newcommand{\figureParameterComparison}{
    \begin{figure}[t]
      \centering
      \includegraphics[width=\columnwidth]{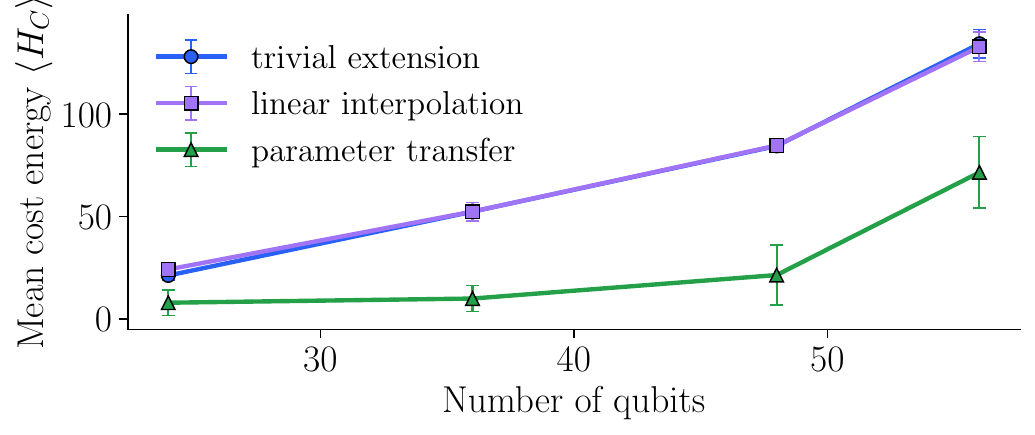}
      \caption{\justifying
      Performance comparison of three QAOA parameter initialization methods for $p=2$ for the MIS on Erdős--Rényi graphs with edge density $q=0.08$: trivial extension, linear interpolation~\cite{zhou2020quantum}, and parameter transfer~\cite{sureshbabu2024parameter}.
      Lower mean cost energies $\langle H_C\rangle$ correspond to better solutions, i.e., higher approximation ratios. 
      Error bars indicate variation over 10 graph instances. 
      }
      \label{fig:parameter_comparison}
    \end{figure}
}

\newcommand{\figurePostselectionCircuitAppendix}{
    \begin{figure}[H]
    \centering
    
\newcommand{\midvdots}{\push{\raisebox{1.5ex}{\(\vdots\)}}}

\begin{quantikz}[row sep=0.4cm, column sep=0.85cm]
&[-0.7cm] 
&[-0.5cm] \gate[4]{\text{QAOA}}\slice[style={gray,dashed}]{} 
& \meter[label style={yshift=-2pt}]{m_1}\slice[style={gray,dashed}]{} 
& \gate{X}\gategroup[4,steps=2,
    style={dashed,rounded corners,draw=black!55,inner xsep=4pt,inner ysep=10pt},
    label style={anchor=south,yshift=-4pt}
]{inversion test} 
&[-0.35cm] \meter[label style={yshift=-2pt}]{m_1'} 
\\
&[-0.25cm] 
&[-0.15cm] 
& \meter[label style={yshift=-2pt}]{m_2} 
& \gate{X} 
&[-0.35cm] \meter[label style={yshift=-2pt}]{m_2'} 
\\
\setwiretype{n}
&[-0.25cm] \midvdots
&[-0.15cm] 
& \midvdots
& \midvdots
&[-0.35cm] \midvdots
\\
&[-0.25cm] 
&[-0.15cm] 
& \meter[label style={yshift=-2pt}]{m_n} 
& \gate{X} 
&[-0.35cm] \meter[label style={yshift=-2pt}]{m_n'} 
\end{quantikz}

    \caption{
    \justifying
    Hardware-based post-selection scheme. 
    After the hardware-aware transpilation of a QAOA circuit, all qubits are measured, giving outcomes $\{m_i\}$, then undergo an inversion test: they are inverted by a calibrated $\pi$ pulse, then measured again, giving $\{m_i'\}$. 
    A shot is kept only if $m_i \neq m_i', \forall i$, and discarded otherwise. 
    The dashed lines mark the barriers separating the three stages.
    }
    \label{fig:postselection_circuit}
    \end{figure}
}

\newcommand{\figurePostselectionAppendix}{
    \begin{figure*}[htbp]
    \centering
    \includegraphics[width=\linewidth]{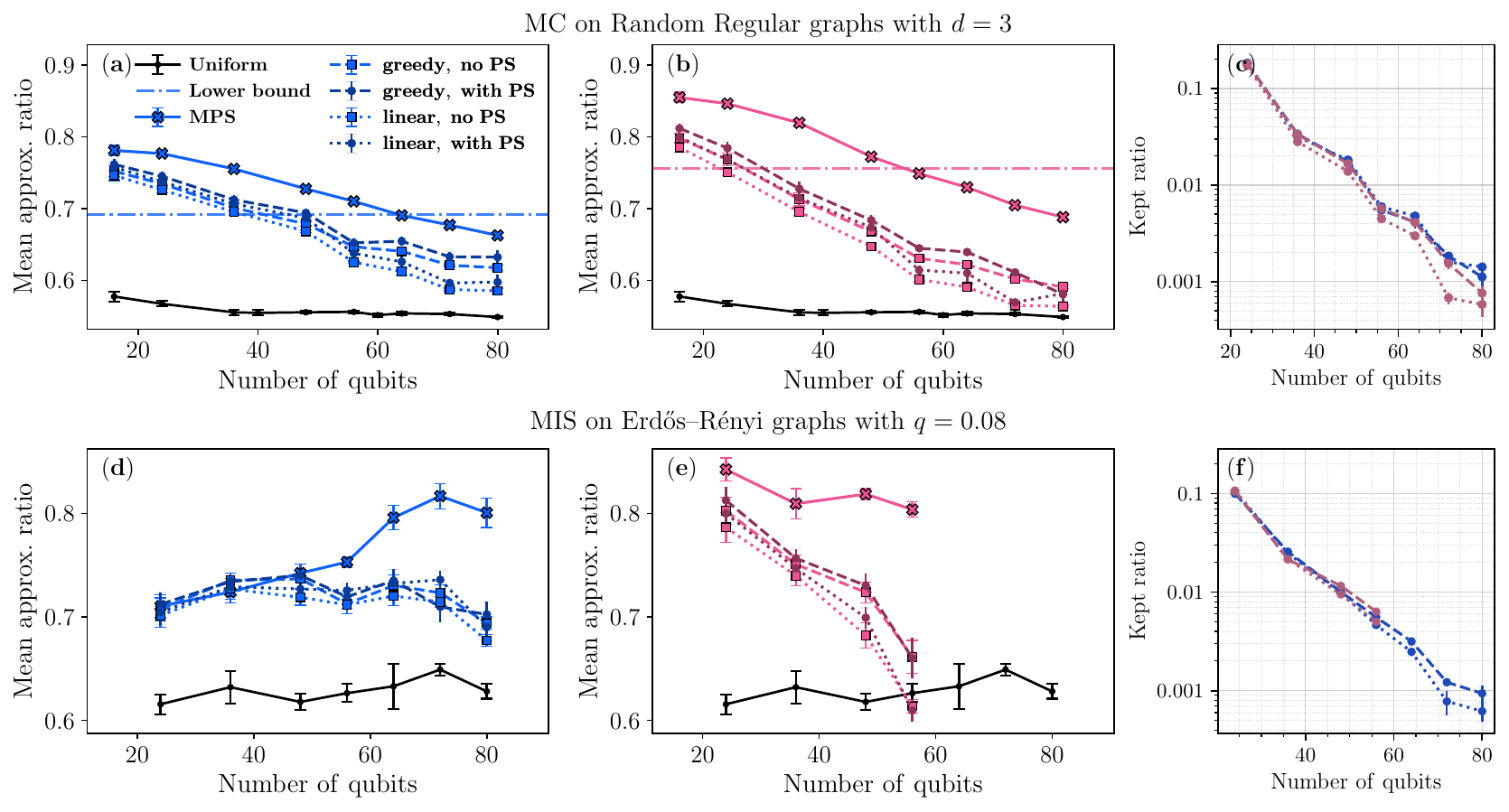} 
    \caption[singlelinecheck=false]{
    \justifying 
    Post-selection (PS) impact on the hardware QAOA circuits with $p=1$ (blue) and $p=2$ (pink) layers, transpiled using \emph{linear} (dotted) and \emph{greedy} (dashed) strategies. 
    Results are averaged over $10$ graph instances, with $10{,}000$ samples per circuit. 
    }
    \label{fig:postselection}
\end{figure*}
}